\documentclass[journal=jpccck,manuscript=article,layout=twocolumn]{achemso} 

\usepackage[version=3]{mhchem} 
\usepackage{graphicx}

\newcommand{\mnras}{Mon.~Not.~R.~Astron.~Soc.~}
\newcommand{\apj}{Astrophys.~J.~}
\newcommand{\aap}{Astron.~Astrophys.~}
\newcommand{\jcp}{J.~Chem.~Phys.}
\newcommand{\apjs}{Astrophys.~J.~Suppl.~}  

\author{Jennifer A. Noble}
\affiliation{CNRS, Aix-Marseille Univ, PIIM, Marseille, France}
\alsoaffiliation{School of Physical Sciences, University of Kent, Canterbury, UK}
\email{jennifer.noble@univ-amu.fr}
\author{Herma M. Cuppen}
\affiliation{Radboud University, Institute for Molecules and Materials, Nijmegen, The Netherlands}
\alsoaffiliation{van ’t Hoff Institute for Molecular Sciences, University of Amsterdam, Amsterdam, The Netherlands}
\author{Stephane Coussan}
\affiliation{CNRS, Aix-Marseille Univ, PIIM, Marseille, France}
\author{Britta Redlich}
\affiliation{FELIX Laboratory, Radboud University, Nijmegen, The Netherlands}
\author{Sergio Ioppolo}
\affiliation{School of Electronic Engineering and Computer Science, Queen Mary University of London, London, UK}
\email{s.ioppolo@qmul.ac.uk}

\title[]{Infrared resonant vibrationally induced restructuring of amorphous solid water} 

\begin{document}




\begin{abstract}

Amorphous solid water (ASW) is abundantly present in the interstellar medium, where it forms a mantle on interstellar dust particles and it is the precursor for cometary ices. In space, ASW acts as substrate for interstellar surface chemistry leading to complex molecules and it is postulated to play a critical role in proton-transfer reactions. Although ASW is widely studied and is generally well characterized by different techniques, energetically-induced structural changes, such as ion, electron and photon irradiation, in these materials are less well understood. Selective pumping of specific infrared (IR) vibrational modes can aid in understanding the role of vibrations in restructuring of hydrogen bonding networks. Here we present the first experimental results on hydrogen bonding changes in ASW induced by the intense, nearly monochromatic mid-IR free-electron laser (FEL) radiation of the FELIX-2 beamline at the FELIX Laboratory at the Radboud University in Nijmegen, the Netherlands. The changes are monitored by reflection-absorption infrared spectroscopy. Upon resonant irradiation, a modification in IR absorption band profile of ASW is observed in agreement with a growing crystalline-like contribution and a decreasing amorphous contribution. This phenomenon saturates within a few minutes of FEL irradiation, modifying upwards of 94~\% of the irradiated ice. The effect is further analysed in terms of hydrogen bonding donors and acceptors and the experiments are complimented with Molecular Dynamics simulations to constrain the effect at the molecular level.

\end{abstract}


\section{Introduction}

Amorphous solid water (ASW) is one of the most widely studied molecular systems due to its ubiquitous presence in the interstellar medium (ISM), on icy bodies in the Solar System, as well as possibly at high altitude in the Earth's and some exoplanets' atmospheres. In these diverse environments, ASW is far from pure, as small molecules such as CO, \ce{CO2}, \ce{CH4}, \ce{O2}, and \ce{N2} accrete onto or form on its surface, where they can undergo surface reactions. Such reactivity contributes to the formation of complex prebiotic species in interstellar space. The ASW is not a mere spectator in this process, but rather actively participates in the surface reactions by, for example, stabilization and proton transfer. The latter strongly depends on the underlying hydrogen bonding network. Studies have shown that, for instance,  H/D exchange through proton exchange has a significantly lower activation energy in ASW than in crystalline ice.\citep{Lamberts:2015} This is most likely due to the higher defect density in the hydrogen bonding network of ASW. 

While water ice is frequently studied,\cite{Bartels-Rausch:2012} the exact nature of its hydrogen bonding network, especially that of amorphous ices, has not been fully revealed.  
In space, energetic processes in interstellar icy grains generate excess vibrational energy that must be dissipated into the ice itself, whether it be cosmic ray (CR) bombardment,\cite{Shingledecker:2019} CR-induced UV photodesorption, or surface chemistry\cite{Islam:2007}. Unfortunately, to date, the specific interaction of vibrational energy with the icy mantle is not addressed in models of these phenomena\cite{Cuppen:2017}, which typically consider bare silicate or carbonaceous grains and restrict energy dissipation concerns to grain heating provoking desorption of low mass adsorbates such as CO\cite{Hasegawa:1993A,Herbst:2006} or to ice grain explosion due to heavy CR-induced spot heating.\cite{Ivlev:2015} Completing the description of energy dissipation processes in icy interstellar dust grains is key to correctly integrating gas-grain processes such as photodesorption, thermal desorption, chemical desorption\cite{Garrod:2007,Dulieu:2013} as well as grain radiation in the infrared (IR) and terahertz (THz) into astrochemical models. The latter processes can potentially cause physico-chemical changes in the ice as molecules in the solid phase -- including water -- absorb intra- and inter-molecular vibrational energy in the IR and THz spectral ranges, respectively.\cite{Ioppolo:2014} A detailed study of energy dissipation in ice grains is also critical in constraining the structure of interstellar icy mantles, where water ice is formed on cold dust grains and expected to be initially amorphous and compacted to some extent by the exothermicity of \ce{H2O} (or other small molecule)\cite{Accolla:2011} formation, but is also exposed to processing events that can alter its structure.

ASW is a metastable state of ice, and radiation could in principle lead to structural modification. To date, the effect of low energy photon irradiation, \emph{i.e.}, in the infrared regime, on the hydrogen bonding network has been scarcely addressed. Only a couple of IR irradiation studies have been performed on crystalline water ice at elevated temperatures, the main purpose of which was to investigate desorption efficiencies and products.\citep{Krasnopoler:1998, Focsa:2003} These studies demonstrated a strong wavelength dependence, with crystalline ice resonantly desorbing upon irradiation with an IR laser in the 3~$\mu$m region. In terms of ASW, some of us have previously studied the porous amorphous solid water (pASW) surface via the ``dangling'' OH stretching modes, \citep{Noble:2014a,Noble:2014b,Coussan:2015} revealing an irreversible restructuring of these surface modes upon selective IR irradiation.

In this work we have systematically irradiated pASW in the MIR (Mid InfraRed) for the first time using the infrared free-electron laser (FEL) FELIX at the FELIX Laboratory. FELIX is a high intensity, pulsed and tunable source, used here to irradiate at discrete, relevant frequencies -- from the OH stretch, to the bending, the libration, and even the interlayer THz modes of water ice -- to investigate irreversible changes induced by the FEL light in interstellar- and Solar System-relevant ices.


\section{Experimental and computational methods}

\subsection{Experiments}

Experiments were carried-out using the Laboratory Ice Surface Astrophysics (LISA) ultrahigh vacuum (UHV) end-station at the FELIX Laboratory, Radboud University, the Netherlands. Details of the experimental setup are given in the ESI. Here we discuss the experimental methodology used during a standard experiment. Briefly, ices were grown in an ultrahigh vacuum chamber with a base pressure of a few 10$^{-9}$~mbar at the lowest substrate temperature of 17~K. Deionised water was purified via multiple freeze-pump-thaw cycles and was dosed onto a gold-coated copper substrate by background deposition (\emph{i.e.} the main chamber is filled with gas from an all-metal leak valve that does not face the substrate). The substrate was held at 17~K during deposition in order to grow pASW or at 150~K to grow Ic. For both ice structures, the substrate was held at 17~K following deposition. For all experiments on pASW, the same (approximate) number of gas phase water molecules were dosed through the dosing line. This was controlled by keeping deposition pressure in the main chamber (10$^{-6}$~mbar) and deposition time (370~s) constant. Ice samples were monitored during deposition and throughout the experiments with Reflection Absorption InfraRed Spectroscopy (RAIRS) using a Bruker Vertex~80v Fourier Transform InfraRed spectrometer with an external mercury cadmium telluride (MCT) detector positioned at a grazing detection angle of 18$^\circ$. A reference background spectrum of 512 co-added scans is collected prior to any other measurement, to remove any signal along the IR beam path that is not coming from the ice sample. MIR spectra of 256 co-added scans were acquired at a resolution of 0.5 cm$^{-1}$ from 4000 -- 600~cm$^{-1}$ before and after deposition. A pASW thickness of $\sim$~0.25~$\mu$m was chosen to ensure photons fully penetrated ices, while the ice had a high enough IR signal-to-noise in absorbance to monitor subtle structural modifications via FTIR spectroscopy.

Ices were then irradiated using FELIX-2 infrared FEL source (\emph{i.e.} macropulses of 5 Hz repetition rate) at frequencies in the MIR (2.7-11.8~$\mu$m) and at an angle with respect to the surface of 54$^\circ$.  All irradiations were carried out for five minutes, to ensure saturation of effects. The spectral FWHM of the FELIX beam was on the order of 0.8~\%~$\delta\lambda/\lambda$ for all wavelengths. The laser macropulse energy at longer wavelengths ($\sim$~50~mJ and $\sim$~150~mJ in the 6~$\mu$m and 11~$\mu$m wavelength regions, respectively) was attenuated to $\sim$~10~mJ to match that in the 3~$\mu$m region. Thus, at all wavelengths the average power was $\sim$~0.03~W, with a laser fluence at the sample of approximately $\sim$~0.2~J/cm$^2$.
We measured the FELIX beam spot size at the substrate surface to be approximately 2~mm in height (diameter), while the FTIR beam height (diameter) is $\sim$~3~mm (with both beams slightly elongated in width due to their incident angles on the substrate). FTIR spectra were acquired before and after irradiation, from which difference spectra were obtained in order to highlight changes in the IR band profiles. All data analysis and spectral manipulation was performed using in-house python scripts.

\subsection{Simulations}

Classical Molecular Dynamics (MD) simulations were performed using the \textsc{LAMMPS} package (version 16/02/16).\citep{Plimpton:1995} Water molecules were treated flexibly using the TIP4P/2005f potential.\citep{Gonzalez:2011B} Three ice samples were used, each containing 2880 molecules: a hexagonal crystalline sample (Ih) which follows the ice rules and two amorphous samples, one with a density of 0.94~g\,cm$^{-2}$ (labelled ASW) and one with a density of 0.78~g\,cm$^{-2}$ (pASW). Both were obtained by simulating the sample at fixed volume at 400~K for 50~ps and then quenching the system to 10~K. The pASW sample has a large pore, but the local density is rather similar to the ASW sample. 
IR spectra and radial distribution functions were obtained for the samples to compare against experiments and to guarantee reasonable starting structures. 
Simulated IR spectra were obtained from the dipole autocorrelation function, taking into account the changing position of the charge site near the oxygen atom. A Blackman window was applied before the Fourier transform.\cite{Harris:1978}

Before and after the heating runs, the number of hydrogen bonds were determined using an in-house python script to determined the hydrogen bonding structure for each water molecule in terms of donor and acceptor, taking into account the periodic boundary conditions. A radical cutoff of 3.5~{\AA} and a radial cutoff of 30~{$^\circ$} were applied.\cite{Humphrey:1996} An additional python script was applied to calculate the tetrahedral order $q$.\cite{Chau:1998,Errington:2001} This order parameter varies between 0 for an ideal gas and 1 for a regular tetrahedron and is determined for each oxygen atom $O_i$ individually through:
\begin{equation}
 q = 1 - \frac{3}{8} \sum_{j=1}^3\sum_{k=j+1}^4 \left(\cos \psi_{jk} + \frac{1}{3}\right)^2,
\end{equation}
where $\psi_{jk}$ is the angle between $O_j$-$O_i$-$O_k$ where $O_j$ and $O_j$ are nearest neighbour oxygen atoms for $O_i$. 

\section{Results and Discussion}
\subsection{Modification induced by irradiation of experimental pASW}
\begin{figure}[ht!]
\includegraphics[width=\columnwidth,trim= 10mm 12mm 05mm 05mm, clip]{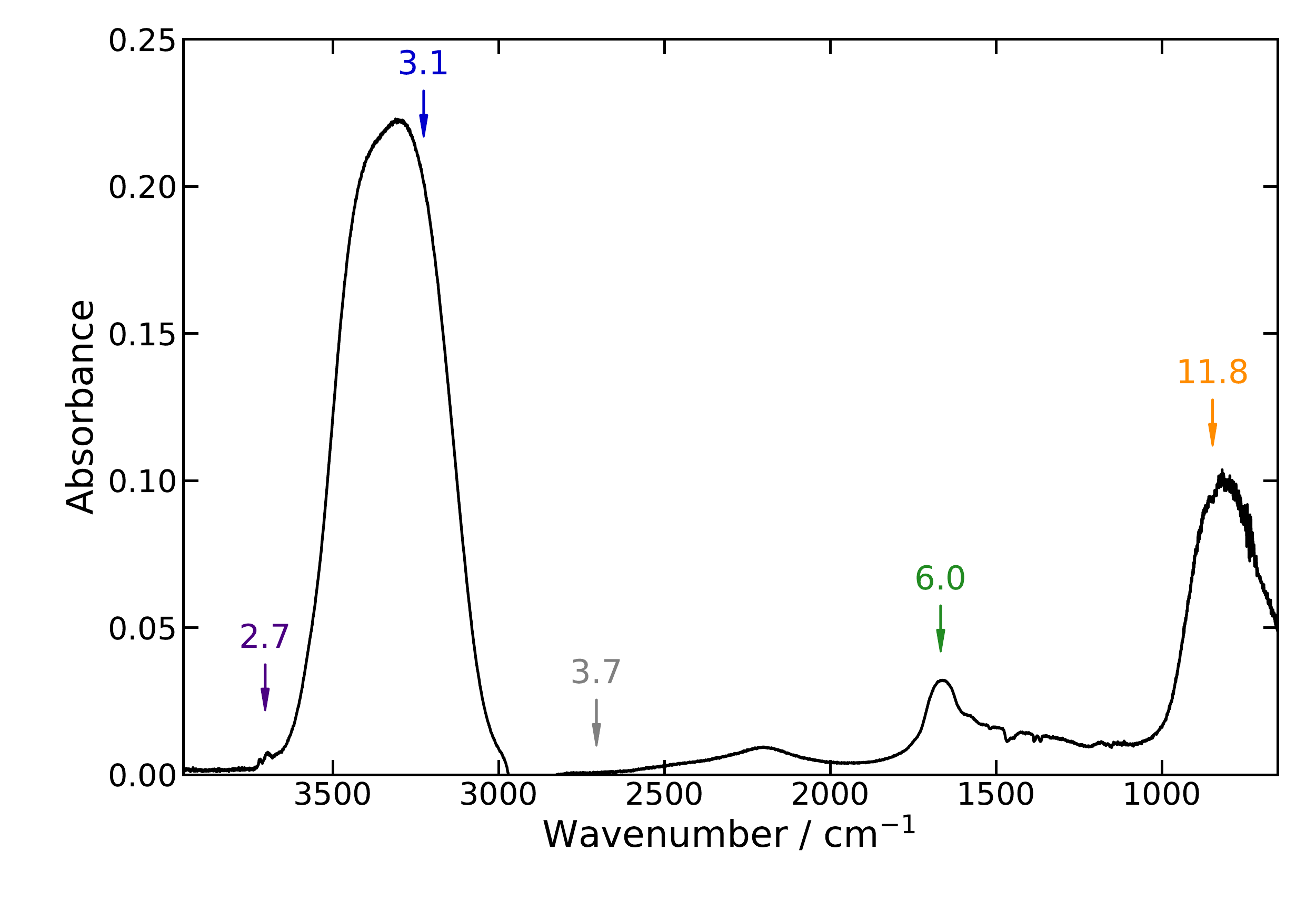}
\caption{Spectrum of deposited pASW ($\sim$~0.25~$\mu$m), marked with the FELIX MIR irradiation frequencies used in this work.}\label{fig:pASW}
\end{figure}

To deposit pASW, ices were grown at a substrate temperature of 17~K onto a gold-coated, optically flat substrate in the LISA end-station at the FELIX Laboratory (see Figure~S1).  For details of the FEL characteristics of the FELIX beam please refer to the `Irradiation  of  the  simulated ices' section. The spectrum of deposited pASW is presented in Figure~\ref{fig:pASW}. The ice was then irradiated using the FELIX-2 ($\sim$~3-45~$\mu$m) infrared FEL source at five different frequencies in the MIR (2.7-11.8~$\mu$m), always for the same duration of time (five minutes) to ensure a full saturation of any possible restructuring effect within the ice, and always on a fresh, unirradiated ice spot. These frequencies are indicated by the coloured arrows in Figure~\ref{fig:pASW}. We will use this colour scheme throughout the paper.

\begin{figure}[ht]
\centering
\includegraphics[width=\columnwidth,trim= 10mm 12mm 05mm 05mm, clip]{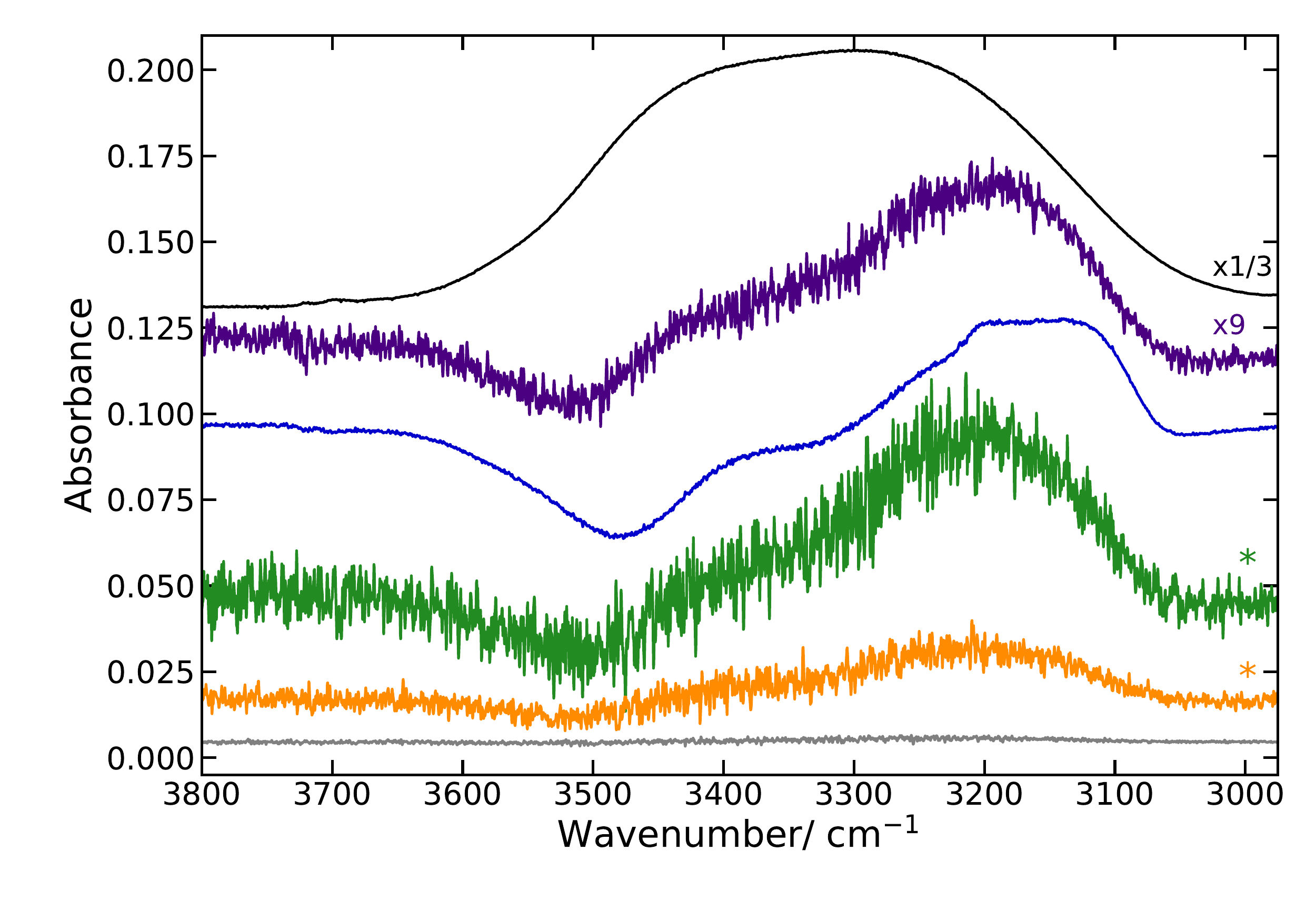}
  \caption{Experimental spectra of pASW before and after irradiation. The pASW signal before irradiation (black line) is divided by three to aid comparison with the difference spectra after irradiation at 2.7~$\mu$m (dangling OH stretching mode, violet), 3.1~$\mu$m (bulk stretching mode, blue), 6.0~$\mu$m (bending mode, green), 11.8~$\mu$m (libration mode, orange), and 3.7~$\mu$m (off-resonance, gray) irradiations. Difference spectra of irradiations at the bending and libration modes are corrected for the band strengths of the irradiated modes (*).\cite{Bouilloud:2015} No band strength is available for the surface modes at 2.7~$\mu$m, and thus for this irradiation (violet) the difference spectrum was multiplied to match the depth of the abundance change after irradiation at 3.1~$\mu$m.}
  \label{fig:thinASWirr_all}
\end{figure}

The spectrum of the OH stretch feature of the deposited pASW is presented in Figure~\ref{fig:thinASWirr_all} (black line), accompanied by the difference spectra observed after irradiation at the 2.7~$\mu$m dangling OH stretch (violet), 3.1~$\mu$m bulk OH stretch (blue), 6.0~$\mu$m bend (green), 11.8~$\mu$m libration (orange) as well as an off-resonance irradiation at 3.7~$\mu$m (gray). The experimental pASW difference spectra all exhibit a similar profile, consisting of loss in the blue wing and gain in the red wing. As discussed later in the text, these data represent evidence of wide-ranging restructuring of an amorphous ice upon selective IR irradiation, as this pattern is indicative of increasing bonding order within the ice. In particular, this is the first time that an ice sample (amorphous or crystalline) has been selectively irradiated at wavelengths corresponding to the bending and libration modes. Here we irradiate the same ice deposited onto a cold gold substrate at different (\emph{i.e.} unirradiated) spots across the wide aforementioned spectral range. Full MIR spectra are presented in Figure~S2, to show the spectral changes observed across the full IR spectral range when ices are exposed to FEL light at selected frequencies.

In Figure~\ref{fig:thinASWirr_all}, the difference spectra for irradiation in the bending and libration modes are corrected for the corresponding band strengths of these modes (1.1~$\times$~10$^{-17}$ and 3.2~$\times$~10$^{-17}$) to allow for a direct visual comparison to the stretching mode irradiation (band strength 2.0~$\times$~10$^{-16}$~cm\,molec$^{-1}$).\cite{Bouilloud:2015} The difference spectrum of the irradiation of the OH-dangling modes is not scaled with its band strength due to the lack of a published band strength corresponding to these surface modes and thus, for ease of presentation, we have chosen to scale it to the same overall depth of absorbance change as the spectrum after irradiation at 3.1~$\mu$m. A qualitative comparison between irradiation at the surface (2.7~$\mu$m) and bulk (3.1~$\mu$m) modes of pASW reveals that, although the overall profile changes are similar, relative peaks are shifted to the blue in the case of irradiation at 2.7~$\mu$m, and a decrease in the OH-dangling modes is more visible when the latter are selectively irradiated. It should be noted that, during the experiment shown in Figure~\ref{fig:thinASWirr_all}, the laser macropulse energy at all wavelengths was attenuated to match that in the 3~$\mu$m region ($\sim$~10~mJ/macropulse). More details can be found in the method section.

The induced modifications in the experimental spectra persist even after irradiation has been switched off, \emph{i.e.} the ice does not revert to its original structure on the timescale of these experiments (24~hours). This indicates that structural change from a metastable to a more stable configuration is responsible for the observed modification. Moreover, the selective injection of vibrational energy into the surface and bulk of pASW induces irreversible modifications of the material despite the fact that at 17~K there is a free energy barrier to form crystalline ice which limits the transformation to the stable structure. It should be noted that, in most cases, we recorded an overall gradual experimental sample temperature increase of approximately 0.02--0.03~K during irradiation, showing that the ice effectively transfers excess heat caused by selective IR irradiation to the gold-coated substrate block. Therefore, we must ask: are structural changes in the ice due to vibrational relaxation through the local H-bonding network or is thermal heating of the gold substrate causing modification in the ice? Figure~\ref{fig:thinASWirr_all} shows that irradiations off-peak do not affect the ice structure. Therefore, the FELIX beam does not directly heat the gold substrate, but instead the ice has to first absorb the laser pulse energy. Moreover, the same figure illustrates that spectral changes upon irradiation at the surface modes (2.7~$\mu$m) are shifted and relatively more intense in the irradiated region compared to those caused by irradiation at the bulk modes (3.1~$\mu$m). These results are in agreement with previous work\citep{Noble:2014a,Noble:2014b,Coussan:2015} and indicate that morphological changes induced by FEL irradiation at certain frequencies are linked to the nature of the vibrational modes affected and their location within the ice.

We verified that changes in the ice are local to the irradiated area by acquiring FTIR spectra in the region immediately above and below the irradiation points on the gold substrate. Hence, the morphological changes are also localised within the area exposed to the laser beam and not a global effect throughout the ice. Moreover, in order to test whether morphological changes in the ice are due to single- or multi-photon processes, we performed a FEL power dependence study on the irradiation of the pASW libration mode (\emph{i.e.} the spectral range where FELIX-2 has nearly a factor of ten more power than in the pASW stretch region) and measured changes at the 3~$\mu$m band (\emph{i.e.} the most sensitive part of the pASW RAIR spectrum to morphological changes). Figure~S3 shows that the increase in intensity of the stretch mode at 3.13~$\mu$m is nearly linearly proportional to the increase in FEL power, in agreement with the hypothesis that changes in the ice are caused by a single-photon process. Figure~\ref{fig:thinASWirr_all} also shows that morphological changes are less prominent when the ice is irradiated at lower frequencies (\emph{e.g.} see stretch vs libration modes). This shows that although single-photon events are less likely to induce changes at lower energies, we still found a linear correlation between laser power and ice changes for the libration mode. However, since local temperature at the irradiated ice spot cannot be measured, we cannot completely rule out a minor contribution to the changes in ice morphology being caused by laser-induced global thermal heating.
This being said, preliminary work on infrared resonant irradiation of CO and CO$_2$ molecules indicates that ices composed of volatile species do not fully desorb upon FELIX-2 irradiation. This means that laser induced heating does not lead to a temperature increase higher than $\sim$~30~K in the exposed ice and hence ice melting effects in the pASW sample are not likely to occur.

In order to determine the nature of the induced modifications in the experimental spectra, it was necessary to fit the difference spectra after irradiation. To this end, synthetic spectra of pASW and cubic crystalline ice (Ic) were generated from optical constants taken from Mastrapa et al.\cite{Mastrapa:2009} using Fresnel equations and experimental parameters, as is standard for thin films, see \emph{e.g.} Horn et al., Teolis et al.\cite{Horn:1995,Teolis:2007} Details of synthetic spectra generation are provided in the ESI, with spectra of pASW and Ic presented in Figure~S4. Based on the data presented in Figure~S4, we can estimate that the original deposited pASW in our experiments was $\sim$~0.24~$\mu$m thick (\emph{i.e.} the closest matching simulated spectrum is 0.24~$\mu$m thick). This is consistent with the thickness assumed from deposition time and pressure in the chamber as well as that derived from the area under the OH stretching band.

\begin{figure}[ht!]
\centering
  \includegraphics[width=\columnwidth, trim= 20mm 87mm 10mm 10mm, clip]{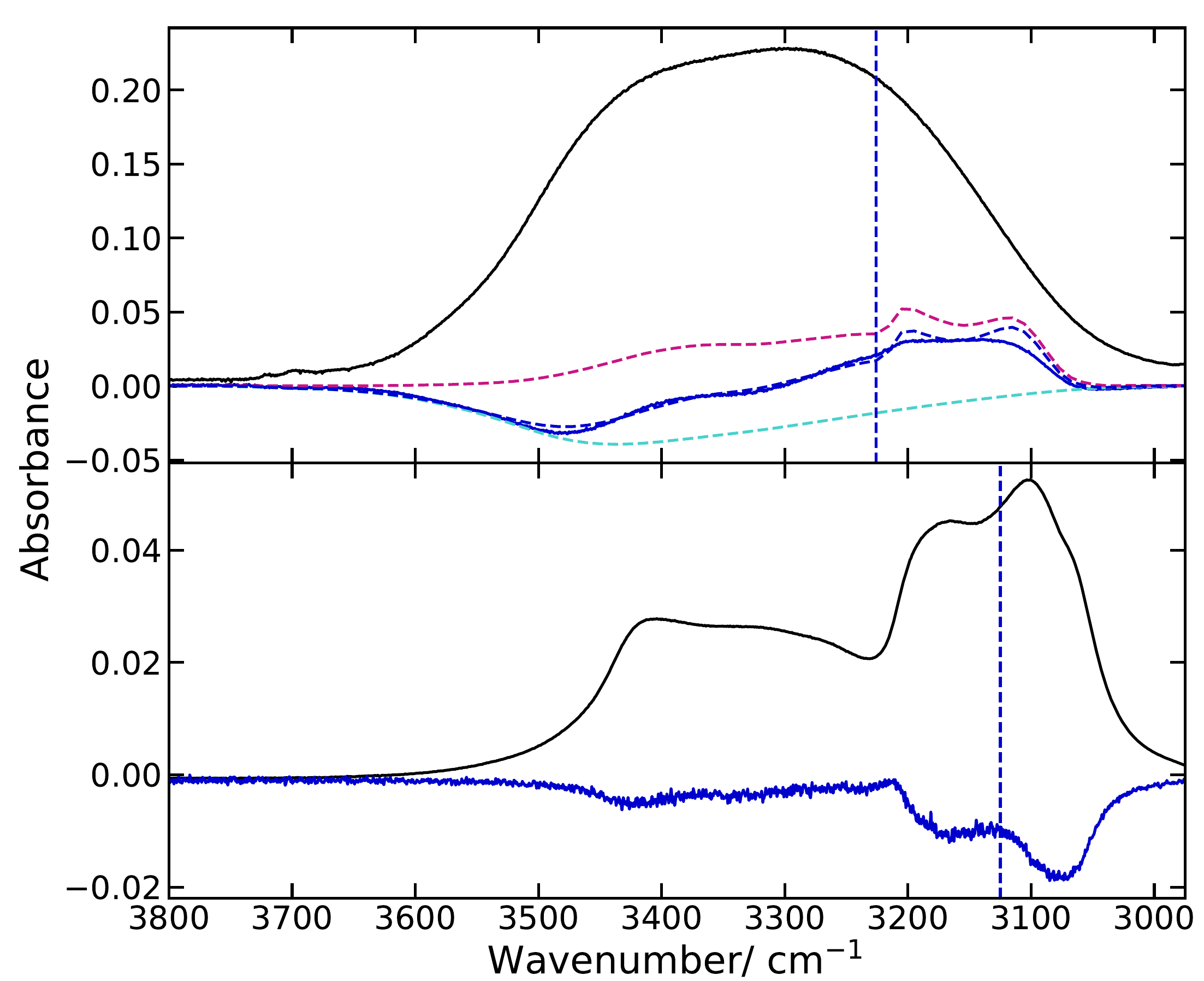}
  \caption{Upper panel: Fit of experimental pASW difference spectrum after irradiation at 3.1~$\mu$m (vertical blue dashed line). The original pASW spectrum (black) is shown to scale. The difference spectrum after irradiation at 3.1~$\mu$m (blue) is fitted by a combination of synthetic spectra for pASW (cyan dashed) and Ic (magenta dashed), resulting in a best fit spectrum (blue dashed). Lower panel: Experimental spectra of Ic before and after irradiation at 3.2~$\mu$m (vertical blue dashed line). The Ic signal before irradiation (black) is divided by ten to aid comparison with the difference spectrum after irradiation at 3.2~$\mu$m (blue).}
  \label{fig:thinASWirr_fit}
\end{figure}

The difference spectrum for the pASW sample irradiated at 3.1~$\mu$m was fitted by a combination of one pASW and one Ic synthetic spectrum using a fitting routine, as presented in the upper panel of Figure~\ref{fig:thinASWirr_fit}. This best fit was achieved by combining a loss of 0.08~$\mu$m pASW (cyan) with a gain of 0.04~$\mu$m Ic (magenta). The combined effect (blue dashed line) reproduces the experimentally-derived difference spectrum (blue continuous line). In particular, the positions of the blue wing oscillator loss and the red wing oscillator gain are well reproduced by the fit, confirming that a significant fraction of the whole pASW sample is modified by irradiation. 

Taking into account the geometry of the overlapping FTIR (3 mm height and an angle of 18$^\circ$ w.r.t. the ice surface) and FEL beams (2 mm height and an angle of 54$^\circ$ w.r.t. the ice surface), this result indicates that upwards of  94~\% of the irradiated pASW is restructured upon resonant IR irradiation. Unfortunately this is not directly visible from the FTIR spectra because the larger size of the FTIR beam compared to the FELIX one causes a dilution effect. The discrepancy between the two ice (pASW and Ic) thickness values (0.08 and 0.04~$\mu$m) is due to multiple factors, chiefly: the relative densities of the two ice phases (with the more compact Ic being intrinsically thinner than pASW); the presumed loss of some small fraction of water molecules due to desorption at grain boundaries; and the inherent uncertainty in using optical constants derived from pure ice phases (pASW at 15~K, and Ic deposited at 150~K and measured at 17~K)\cite{Mastrapa:2009} to fit an effect in what is likely -- after irradiation -- an ice of mixed phase. This latter is particularly hard to constrain. How can one quantify the degree of ``crystallisation'' that has occurred in the ice when one cannot precisely simulate the mixed nature of the resultant ice? Ice mixtures have previously been expressed in terms of the Maxwell-Garnett mixing rule,\cite{Mukai:1986, Bossa:2014} wherein optical constants from two pure ices are combined to describe the optical properties of the mixed ice. In this case, however, the nature of the mixing state of our ice is undefined, as variables such as the size of localised crystalline nuclei, the extent of pore collapse, and the degree of reorganisation cannot be determined. The penetration depth of photons in pASW at 3~$\mu$m is $\sim$~0.5~$\mu$m,\cite{Mastrapa:2009} and thus, because the thickness of our pASW was always $<$~0.3~$\mu$m, we can assume that roughly the whole sample was irradiated. As such, at this time we choose to perform a simple fitting of the difference spectra with pure simulated pASW and Ic ices in order to distinguish the contributions of unmodified amorphous ice from restructured areas.

Due to the presence of crystalline-like features in the spectra of our irradiated pASW samples, we performed a control experiment to irradiate Ic at 17~K. A similar experiment has been previously performed by two groups for the OH stretching mode of crystalline ices.\cite{Krasnopoler:1998,Focsa:2003} The results of our irradiation of an Ic sample at 3.2~$\mu$m are presented in the lower panel of Figure~\ref{fig:thinASWirr_fit}. In contrast to our pASW, where both loss and gain were observed in the difference spectrum, in the case of Ic only loss of ice is observed. In fact, the profile of the difference spectrum almost exactly mirrors that of the original deposited ice, confirming that the Ic undergoes an approximately pure desorption upon resonant IR irradiation. As for pASW, no modification of the spectrum is observed upon off-resonance irradiation. Based on the observed 3~$\mu$m band loss of spectral absorption, we estimate that roughly 15~\% of the ice desorbs upon FEL exposure in the irradiated spot area. In previous studies on Ic (\emph{i.e.} ices deposited at 140/$>$150~K and irradiated at 110/100~K),\citep{Krasnopoler:1998, Focsa:2003} where \ce{H2O} desorption was measured in terms of the number of water molecules desorbing from the surface as a function of irradiation frequency, the derived desorption profile was observed to reproduce, respectively, a liquid water and a crystalline ice spectral profile. When we irradiate crystalline ice (Fig.~\ref{fig:thinASWirr_fit}, Lower Panel), we obtain the same results as \citet{Focsa:2003}, \emph{i.e.} desorption from the ice phase, rather than, as postulated by Krasnopoler \& George,\cite{Krasnopoler:1998} desorption from a pre-melted liquid water layer. Differences between our results and those from Krasnopoler \& George,\cite{Krasnopoler:1998} can be explained by the choice of different experimental conditions, \emph{e.g.} laser power and focusing, ice deposition method, and our irradiation at low temperature ($\sim$~17 K). We would like to point out that the advantage of the methodology in our study compared to the previous investigations is that, in using FTIR spectroscopy to trace the whole ice sample, we have an overview of the impact of irradiation at a single frequency on all frequencies simultaneously, rather than simply measuring the global desorption effect induced at a single frequency. We reaffirm that the profile of the desorbing crystalline ice (\emph{i.e.} the difference spectrum) almost exactly matches the profile of the deposited ice and no oscillator-specific modification is observed. This is also confirmed by irradiating Ic at different frequencies in the 3~$\mu$m spectral region (not shown here). Thus, we conclude that irradiation of Ic leads to surface desorption and not to reorganisation of oscillators. This is in line with our expectations, since deposited Ic ice is already in its stable form under these conditions.

\subsection{Irradiation of the simulated ices}

Classical Molecular Dynamics (MD) simulations allow one to simulate irradiation experiments by employing an oscillating electric field. In our simulations we see that vibrations in the molecules are resonantly excited. Since, experimentally, the structural changes have been observed to build up over a timescales of several minutes and since structural changes have been observed, we believe that classical simulations that allow for longer timescales and larger system sizes are more appropriate here than quantum chemical methods. For details on the MD simulations please refer to the method section. A discussion on quantum effects can be found in the ESI.

\begin{figure}
 \centering
 \includegraphics[width=\columnwidth]{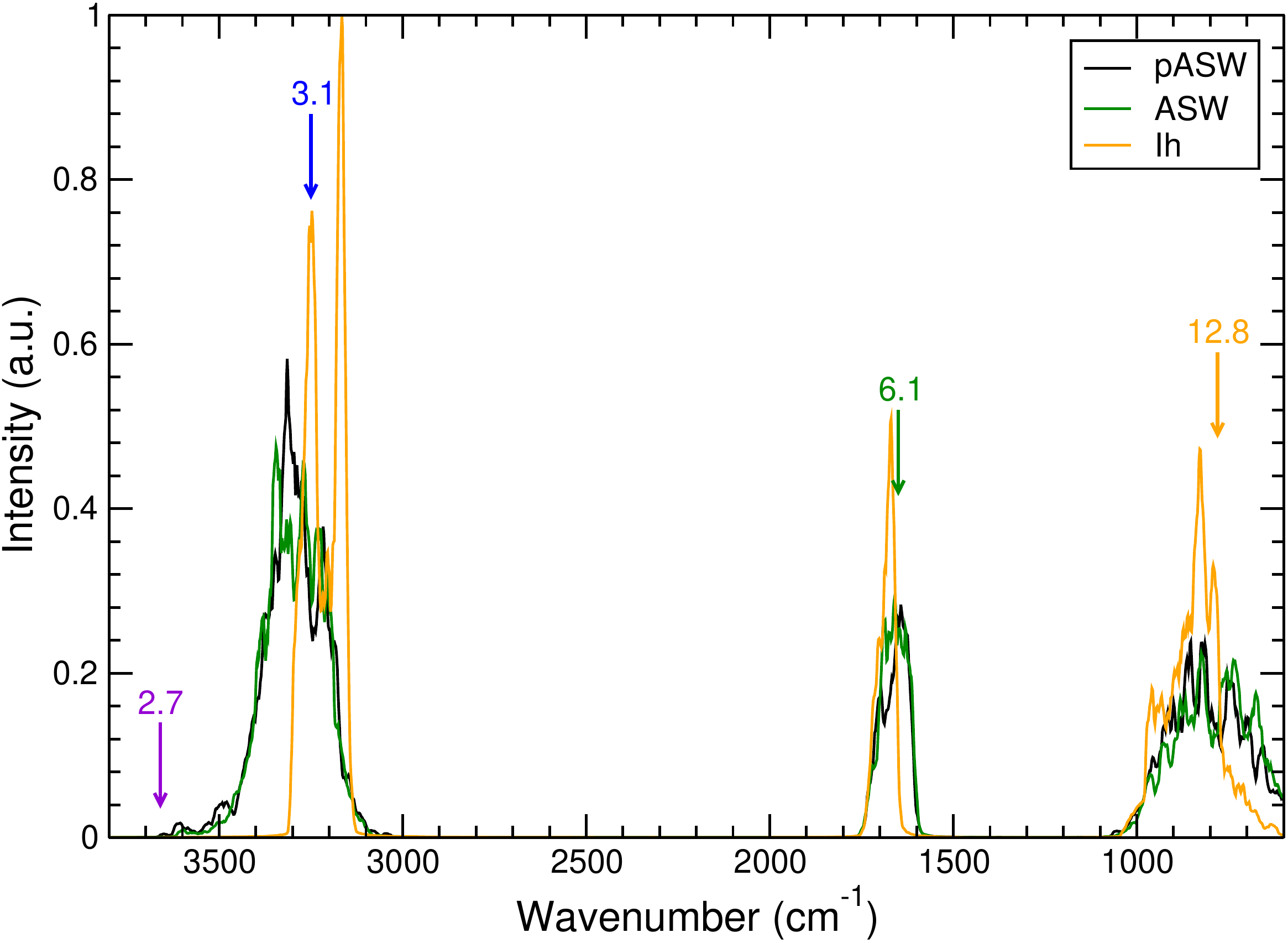}
 \caption{Simulated IR spectra of the pASW, ASW and Ih samples, marked with the frequencies of the oscillating electric field.}
 \label{fig:simspec}
\end{figure}

We use three ice samples, each containing 2880 molecules: a hexagonal crystalline sample (Ih) as a reference system, and two amorphous samples with different total densities (ASW, pASW). The local density between these two latter samples is very similar, but the pASW sample contains a large pore with a pore wall (\emph{i.e.} a higher surface to bulk ratio).
Simulated IR spectra of the three samples are presented in Figure~\ref{fig:simspec}. The crystalline features are much more narrow and peaked, in accordance with experiments. The pASW sample shows small dangling OH features between 3600 and 3700~cm$^{-1}$, indicating the contribution of molecules at the surface of the pore. In the remainder, we will focus on the pASW results and we will only report deviations from this for the other two samples.

As mentioned above, irradiation of the ice samples was achieved by the application of an oscillating electric field along the $z$-direction, with a maximum amplitude of 15~mV/\AA.
For the FEL beam settings chosen in this work, FELIX-2 consists of a train of 10$^{4}$ (1~GHz repetition rate) micropulses of typically 0.3--5~ps, which are 1~ns apart, and form 10~$\mu$s macropulses spaced 200~ms (5~Hz repetition rate) apart. While the energy of a single micropulse is $<$40~$\mu$J, a macropulse has a typical energy $<$200~mJ in the 3--45~$\mu$m frequency range. Our MD simulations show that during the time between micropulses (1~ns), the ice can relax again. We expect cooling through the thermostat to be the most important contribution. It is hard to determine the local cooling power at the experimentally irradiated spot, but we estimate 10~W to be a very conservative upper limit. Considering the volume of the irradiated ice, the thermal energy taken out of the system is  $3$~kJ~m$^{-3}$ in 1~ns. Unfortunately, these long relaxation times can not be reached in the molecular dynamics runs. Instead, we choose the following setup: fifteen cycles are performed, consisting of 2~ps perturbation and 18~ps relaxation, where the perturbation is either IR irradiation using an oscillating electric field or heating by a thermostat. Four IR frequency are used: 12.8, 6.0, 3.1, and 2.7~$\mu$m. During the 18~ps of relaxation time, only part of the cell is thermostated in order to reduce the cooling power. Kinetic energy taken from the system now corresponds to $3\times10^3$~kJ~m$^{-3}$ in 18~ps and is hence still too much relative to the assumed cooling power in the experiments. At the end of the 15 cycles, the thermostat is finally switched on in the full cell.

\begin{figure*}[ht]
\centering
  \includegraphics[width=0.8\linewidth]{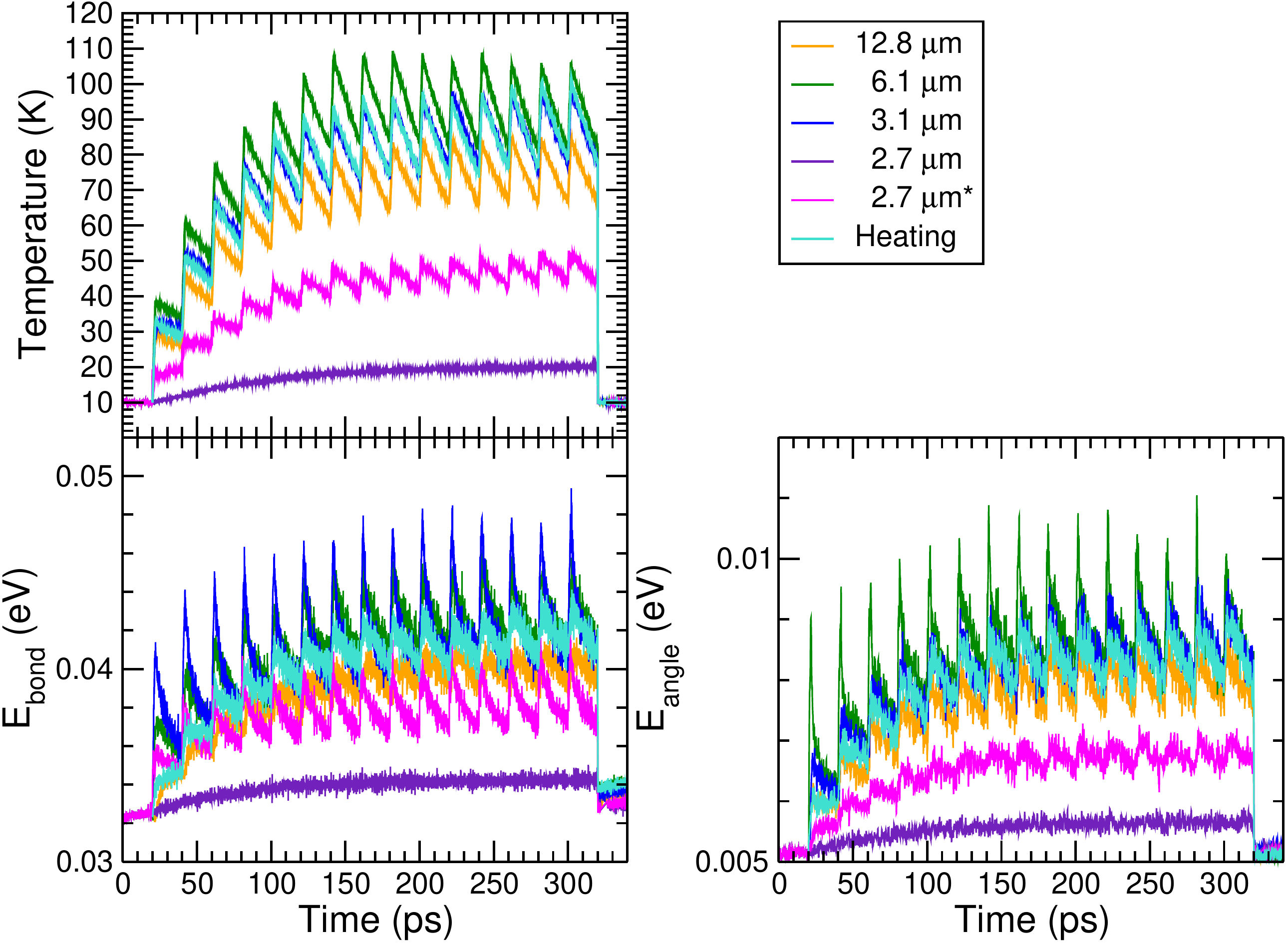}
  \caption{Irradiation effects in simulated pASW in terms of temperature, energy in the bonds and energy in the H-O-H angles of the water molecules as a function of time. The pASW sample is exposed to an oscillating electric field for 2~ps, followed by 18~ps without electric field. Only 10~\% of the water molecules is thermostated to 10~K. This is 15 times repeated, after which the full cell is thermostated. The electric field clearly induces an increase in temperature.  Irradiation at 2.7~$\mu$m was performed at two values of 15~mV/{\AA} (standerd value (violet curve) and 150~mV/{\AA} (pink curve). The cyan curve was achieved by adopting a heating profile to match the profile of 3.1~$\mu$m irradiation (blue curve).}
  \label{fig:simulated_irr}
\end{figure*}

Figure~\ref{fig:simulated_irr} shows the temperature profile, the energy in bonds, and that in angles during such a simulation of 15 perturbation cycles. Irradiation at 2.7~$\mu$m was performed at two values of 15~mV/{\AA} (violet curve) and 150~mV/{\AA} (pink curve). Clear heating-cooling cycles can be observed for all frequencies except 2.7~$\mu$m at low intensity (violet). This is probably since there are only a few surface water molecules available that can be resonantly excited at this frequency. For the heating curve (cyan), a heating profile was adopted to match the profile of 3.1~$\mu$m irradiation (blue curve). The resulting ice might help us distinguish heating from irradiation effects. 

Together with the temperature increase, an increase in both bond and angle energy can be observed during irradiation. This can be partly explained by the temperature increase of the sample (see heating curve, cyan). However, the extent of the effect also depends on the mode that is excited. Irradiation at OH-stretch frequencies (3.1 and 2.7~$\mu$m) results in a stronger effect on the binding energy, whereas at the bending frequency (6.1~$\mu$m) the effect is strongest in the angle energy. 
For the crystalline sample and the ASW sample without a pore, very similar heating profiles are obtained.

\begin{figure}[ht]
\centering
  \includegraphics[width=\columnwidth]{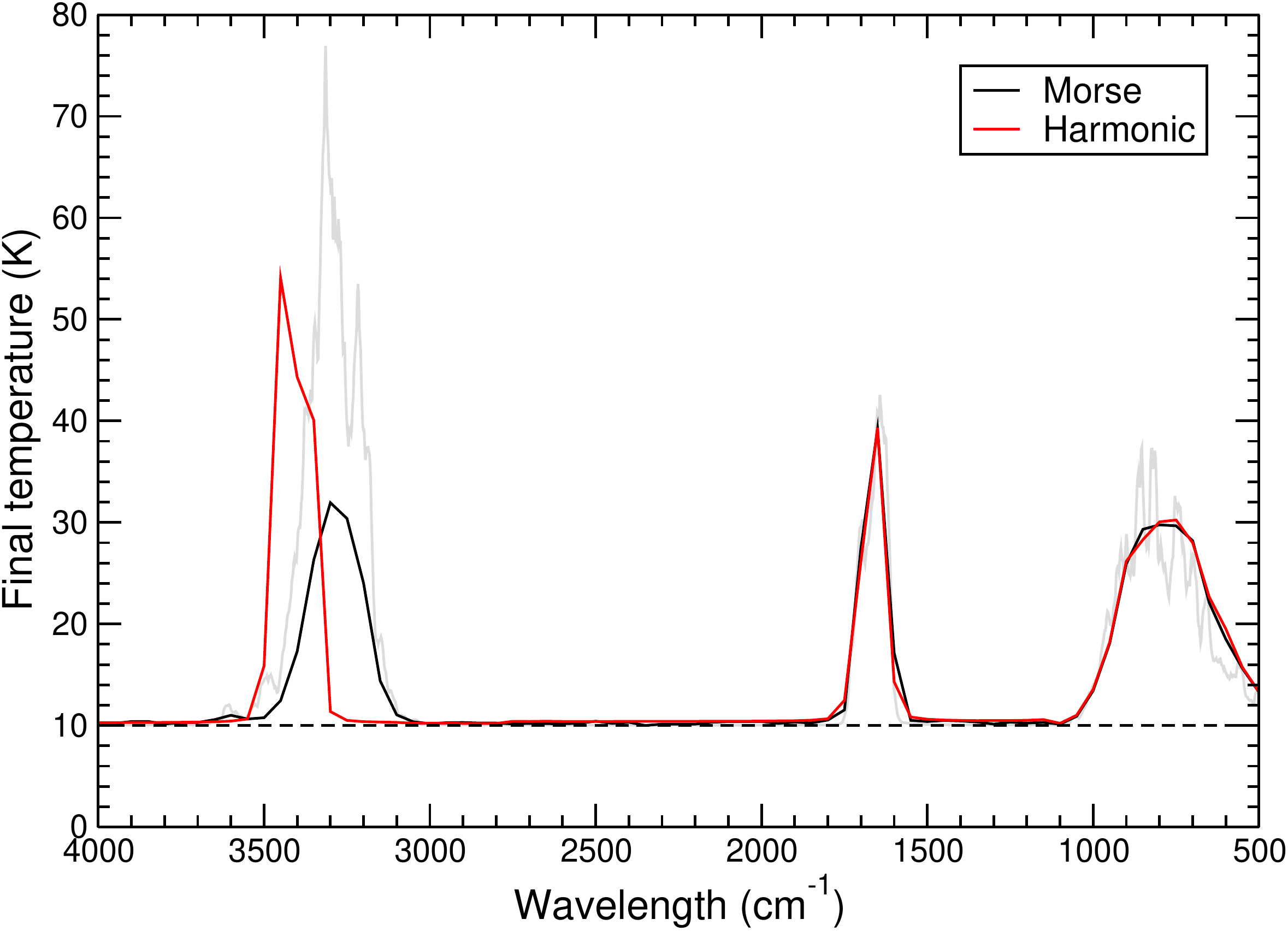}
  \caption{Final temperature reached after 2~ps of exposure as a function of the frequency used. The initial temperature is 10~K (dashed line). The IR spectrum calculated for this sample is plotted in gray to guide the eye. It is scaled and shifted to match the temperature profile at the bending and libration frequencies. }
  \label{fig:temp}
\end{figure}

The heating is directly related to the resonant vibrational excitation, as illustrated in Fig.~\ref{fig:temp}. This shows the temperature that is reached after the first perturbation cycle for different frequencies. The starting temperature is 10~K (dashed line). The IR spectrum presented in Fig.~\ref{fig:simspec} for this samples is replotted in gray. It is scaled and shifted to match the temperature profile at the libration frequencies. Off-resonance, the temperature  remains 10~K, whereas after on-resonance irradiation in the bending and libration frequencies, the temperature profile closely resembles the IR spectrum. For the OH stretch frequencies, the profile is also reproduced, but the highest temperature that is reached is much lower than can be expected based on the IR adsorption strength.  We believe that is due to the anharmonic nature of the OH stretch interaction. Excitation of an anharmonic Morse potential in a classical description results in a vibration with larger amplitude but at slightly different frequency. The excited oscillator is no longer resonant with the field and the effect stops. The bending interaction is much more harmonic and the excited oscillator vibrates with larger amplitude and the same frequency and is hence available to become even more excited. The red curve in Fig.~\ref{fig:temp} uses an harmonic potential for both the angle and the bond, with the resulting heating profile more closely reproducing the IR spectrum for all excitation modes.

How does this relate to the experiment situation, where the effect of irradiation at bending and libration is found to be much lower than at stretch frequencies? Partly, this can be explained by the difference in spectral intensities between simulations and experiments. The simulated spectra in Fig.~\ref{fig:simspec} also exhibit a much higher intensity ratio between the bend and stretch features as compared to the experimental spectrum.  Moreover, we expect the effect of the anharmonicity to be much less severe in the experimental conditions. The pulse in the simulations is at a single frequency and has no intrinsic width. Excited oscillators that shift in frequency are hence immediately outside the range of the pulse and cannot be further excited. In the experimental situation, excited oscillators will likely not shift completely off resonance due to the intrinsic spectral width of FELIX and hence excitation of the bending mode will be more efficient than in the simulations. Moreover, because of the long "off-time" between pulses, most oscillators will have fallen back to the ground state, releasing the energy to the ice, and are thus available again for pumping.

\begin{figure}
    \centering
    \includegraphics[width=\columnwidth]{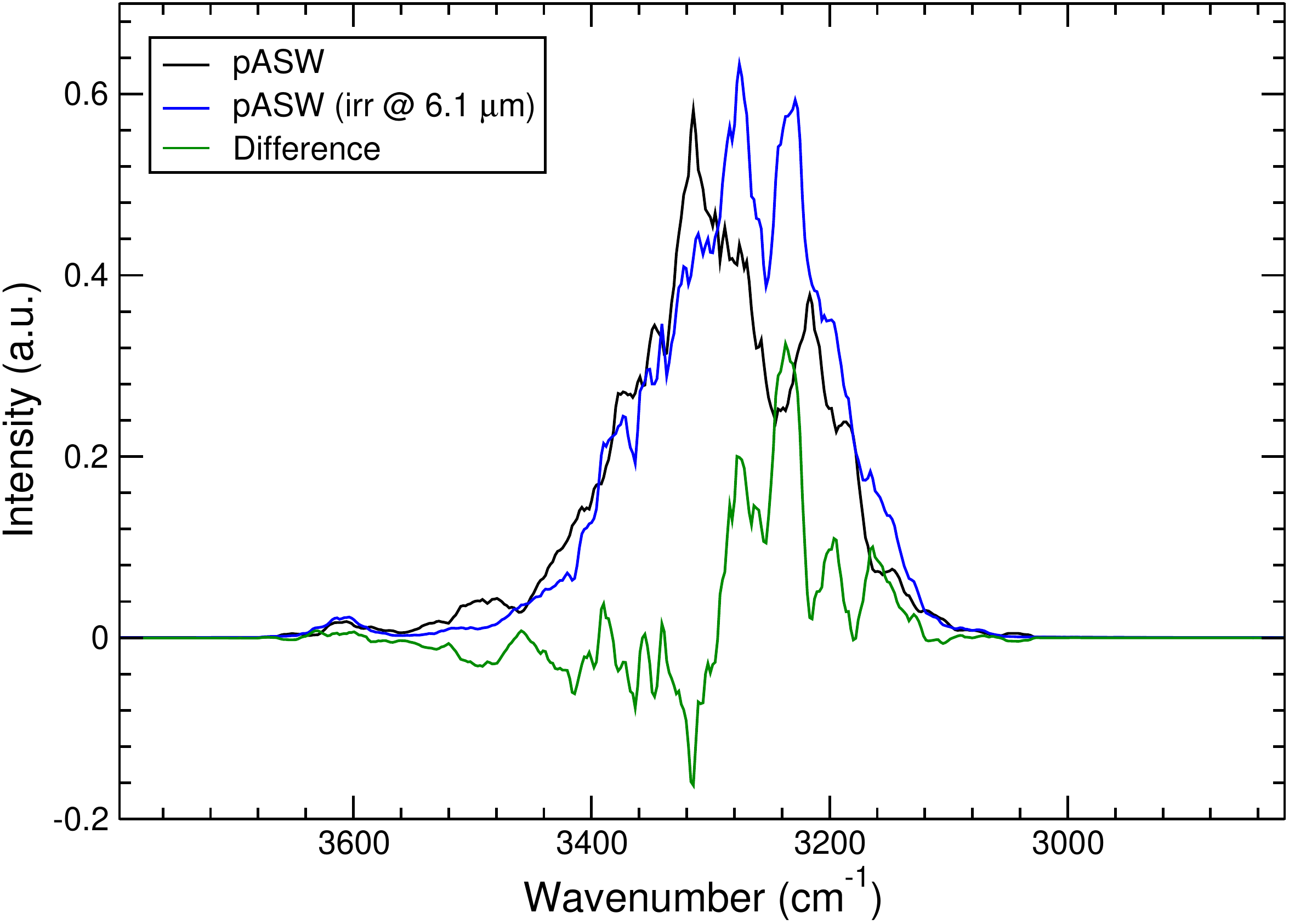}
    \caption{Simulated spectra of pASW before and after irradiation at 6.1~$\mu$m and the difference spectrum over the same frequency range as Fig.~\ref{fig:thinASWirr_all}. }
    \label{fig:MD_spec_diff}
\end{figure}

Figure~\ref{fig:MD_spec_diff} shows the simulated difference spectrum for irradiation of the pASW sample at 6.1~$\mu$m. The spectrum (green trace) clearly exhibits a loss in the blue wing and gain in the red wing, consistent with the experimental finding. In the simulations, these difference features are only observed if the system is allowed to heat-up during irradiation. If the full system is thermostated -- and the whole system remains at 10~K -- no difference in the spectrum is observed after irradiation. This further strengthens our conclusion that the observed differences are due to ice restructuring. Since the cooling profile in the simulations is different from the experimental cooling profile, we expect our simulation results to be a lower limit of the effect, since the experimental ices remain at a higher (local) temperature for a much longer time than in the simulations. Indeed, the overall experimental sample temperature  increased very slightly during irradiation by approximately 0.02--0.03~K, which is reasonable since only roughly 0.4~\% of the total icy substrate surface is irradiated, the Silicon diode (\emph{i.e.} thermal sensor) is at the bottom of the substrate block (\emph{i.e.} $>$15~mm from the irradiation spot), and the time resolution of the temperature measurement is not sufficient to register peak temperatures corresponding to FELIX pulses.

\subsection{Nature of the induced modifications}\label{sec:simulatedspec}

Having confirmed that vibrational excitation induces restructuring of an amorphous ice, we now turn to a more microscopic analysis of the effect of irradiation on localised ice structure. To this end, we analyse experimental spectra in terms of the oscillator families identifiable within the ice structure. There is a rich history of oscillator identification in ices via combined experimental (Raman, IR, SFG) spectroscopy and theoretical simulations of clusters and ices.  A combination of five known oscillator modes in the bulk of the ice spectrum (between $\sim$~3050--3450~cm$^{-1}$) plus three surface-specific modes (two dH at 3720 and 3698~cm$^{-1}$, and the dO at 3549~cm$^{-1}$/s4 at 3503~cm$^{-1}$) were identified from literature data.\cite{Suzuki:2000, Rowland:1991, Buch:1991, Noble:2014a, Smit:2017} For all experimental spectra (original spectra, irradiated spectra, and difference spectra) the same combination of these eight Gaussian functions (G1 -- G8) was fitted to each spectrum, with identical constraints placed on peak position and full width half maximum (FWHM). The areas under these oscillator classes were calculated for all irradiation frequencies. The fitting method is illustrated in the ESI.

\begin{figure}[ht]
\centering
  \includegraphics[width=\columnwidth]{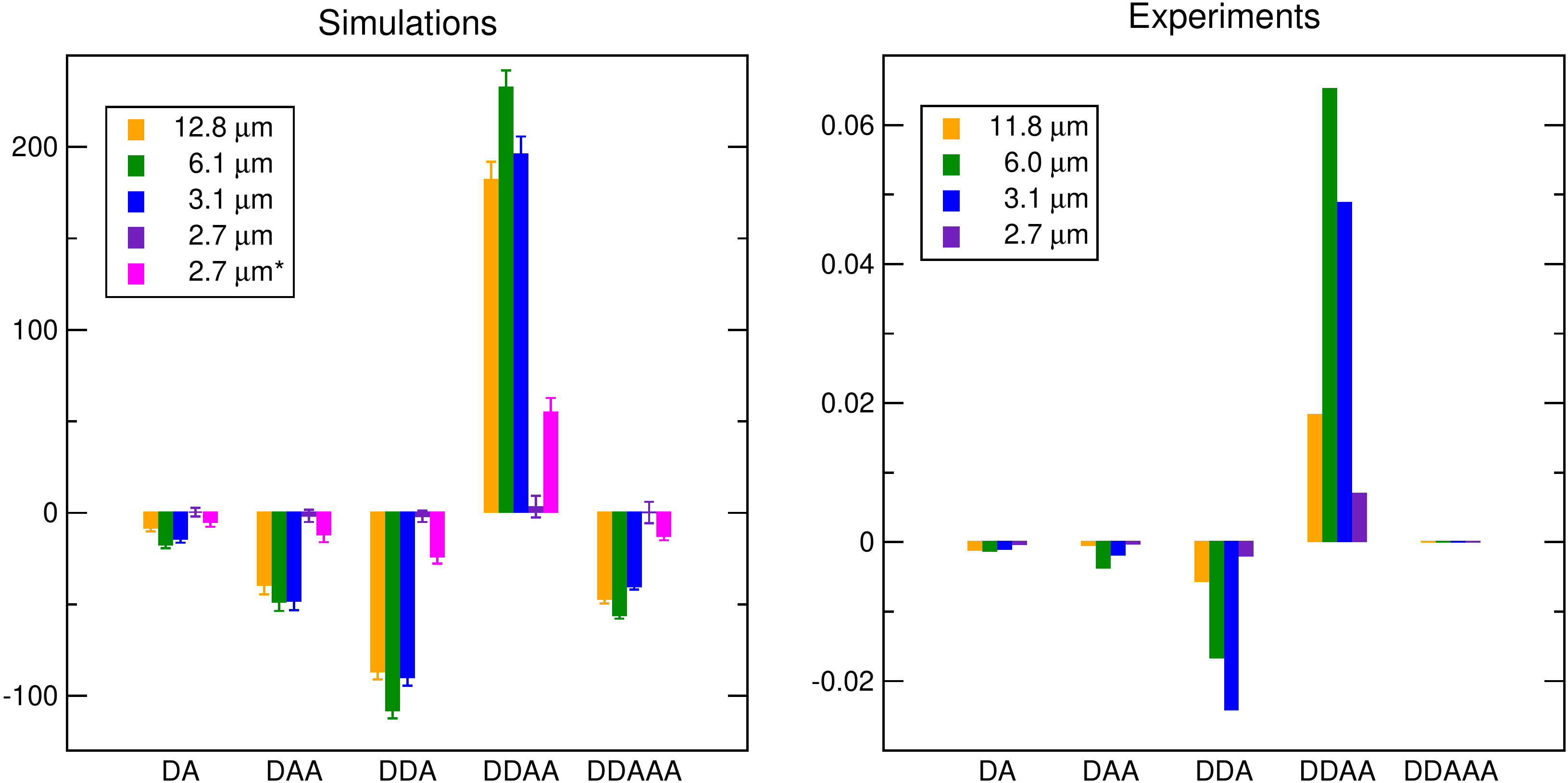}
  \caption{Oscillators impacted by irradiation. Experimental (right panel) compared to simulated (left panel). For the experimentally derived oscillators, the values at 11.8 and 6.0~$\mu$m are corrected for band strength to allow direct comparison to the data at 3.1~$\mu$m. As noted in the text, no band strength is available for the oscillators at 2.7~$\mu$m and thus no correction is made. For the simulated results, the average fraction of molecules in a specific hydrogen bonding configuration during the first 20~ps of the simulation (Fig.~\ref{fig:simulated_irr}) is subtracted from the average fraction during the last 20~ps. The error bar indicates the standard deviation of this fraction.  }
  \label{fig:oscillators}
\end{figure}

Most oscillator classes have also been classified in terms of the local hydrogen bonding structure. These can be directly compared to the molecular data available from the Molecular Dynamics simulations. We compare the modification of oscillators after irradiation for the experimental and simulated ices in Figure~\ref{fig:oscillators}. The differences in areas of these features in the irradiated ices compared to the original unirradiated ice are plotted in the right panel of Figure~\ref{fig:oscillators} for experimental data and the left panel for simulated ices. The oscillator classes are labelled depending on their bonding order within the ice network, as follows: D stands for ``donor'' and A for ``acceptor''. Thus, the least bound molecules (dH feature at 3720~cm$^{-1}$) are labelled DA, followed by the triply bound DAA and DDA (dH at 3698~cm$^{-1}$ and dO at 3549~cm$^{-1}$), and the fully tetrahedrally bound species in the bulk of the ice (DDAA).

In the case of the simulated ice, it was possible to identify quintuply bound molecules (DDAAA). These oscillators have not been assigned to a particular vibrational frequency in experimental spectra, although they most likely contribute to the ``defect'' signatures in oscillators at G8 ($\sim$~3090~cm$^{-1}$). Again, very similar effects were obtained for the other amorphous samples in the simulations. 

\begin{figure}[ht]
\centering
  \includegraphics[width=\columnwidth]{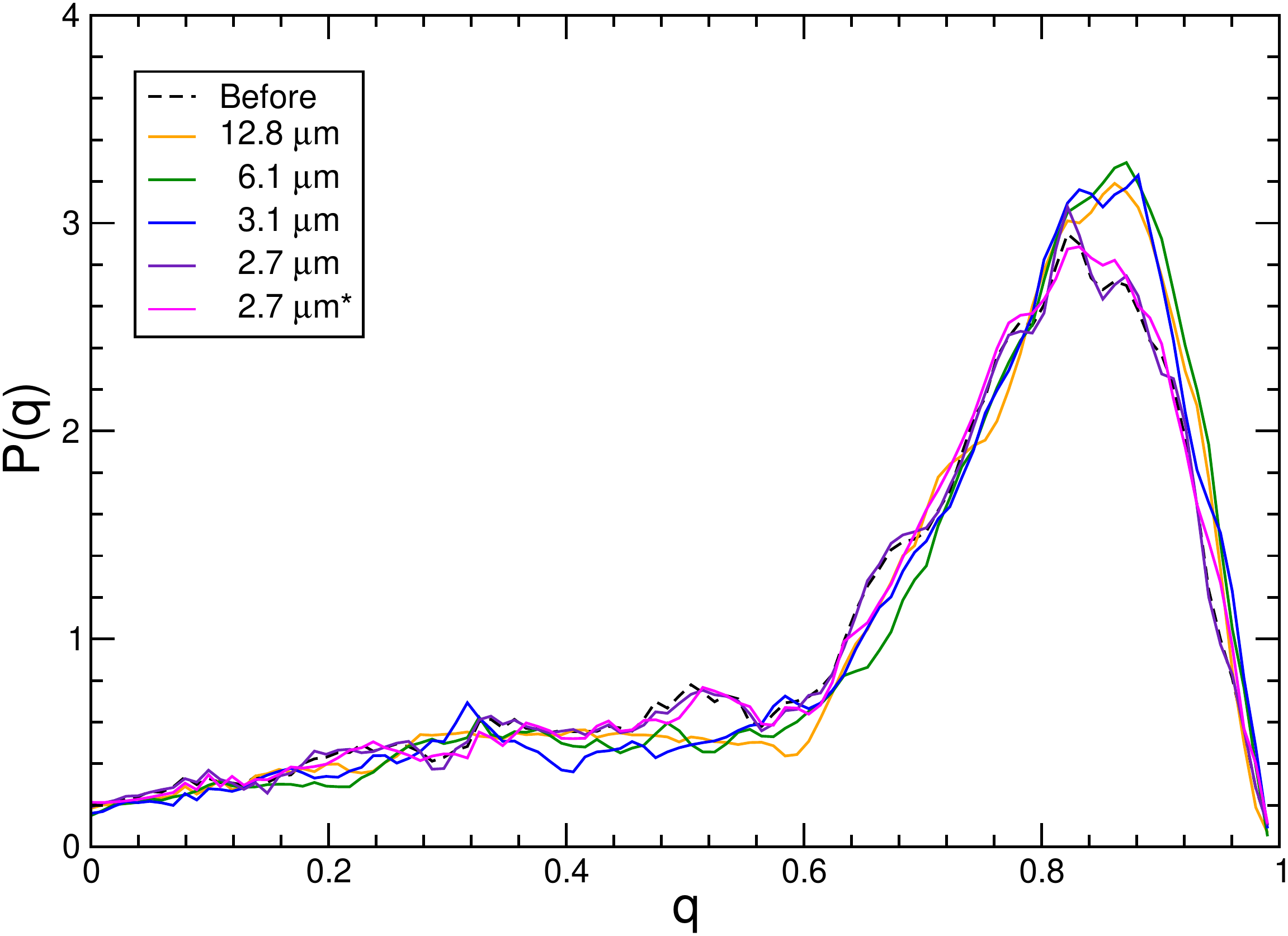}
  \caption{Distribution of tetrahedral order parameter $q$ before (first 20~ps in Fig.~\ref{fig:simulated_irr})  and after irradiation at various frequencies (last 20~ps). A $q$ of one corresponds to a perfect crystalline surrounding.}
  \label{fig:q}
\end{figure}

In a crystalline ice, each water molecule is surrounding by four other water molecules following a regular tetrahedral. So far, for the simulated ices, we have only counted the number of hydrogen bonds per water molecule, without taking into account their relative directions. The tetrahedral order parameter $q$ accounts for this factor. It reaches a value of one for a perfect tetrahedral surrounding, like in crystalline, while a liquid exhibits a broad distribution between zero and one. Figure~\ref{fig:q} shows the obtained distribution of $q$ for our simulated ices before and after irradiation. Except for irradiation at $2.7$~$\mu$m, a clear increase at $q = 0.9$ can be observed, indicating a more crystalline-like structure at the expense of a less well-defined local structure ($q=0.5$).

For the simulated crystalline ice sample, no structural changes were observed after irradiation. In the crystalline ice, all molecules were already in DDAA, with $q=1$, at the start of the simulations. The simulations hence confirm our conclusions from the experiments that IR irradiation induces relaxation to a more stable configuration but does not induce the formation of defects.

Some perspective on our amorphous solid water irradiation study can be gained by considering the wealth of literature on liquid water irradiation and vibrational relaxation. In recent years, a combination of simulations and pump-probe experiments has revealed the heterogeneity of H-bond strengths throughout the water structure, within which the intermolecular interactions are delocalised. From simulations by Imoto et al.\cite{Imoto:2015}, the coupling of vibrational modes is revealed to be critical in explaining the various relaxation timescales in a water sample. Energy is transferred from the OH stretch to the first hydration shell in 50~fs, then to the bending mode in 230~fs, and from there coupling to the libration and translational motions result in a relaxation on the timescale of $\sim$~1~ps. They also find that four- and five-coordinated water molecules relax energy more quickly than the triply-bound surface molecules. This was confirmed by an experimental sum-frequency generation spectroscopy study, where characteristic vibration relaxation timescales at the water-air interface were found to be longer than in the water bulk.\cite{Post:2015} The relaxation time was also found to increase for energy injected into the OH stretching mode at the blue end, compared to relaxation times for energy in the red end of the band. This further suggests that vibrational energy remains trapped at the surface longer than in the bulk. We expect that similar effects occur in pASW ice in short timescales. However an experimental investigation of such effects requires a time resolution beyond the capabilities of the current setup. Hence future experimental studies are needed to time resolve the energy relaxation in pASW samples at low temperatures.

\section{Conclusions}

This study focuses on the selective MIR irradiation of pASW at 17~K by means of the intense, tunable radiation of FELIX-2 free-electron laser at the FELIX Laboratory. Experiments are carried-out by using the LISA end-station, where ice samples are grown and monitored $in~situ$ by Fourier Transform Reflection-Absorption Infrared spectroscopy in the MIR range. MD simulations are here used to constrain the effect at the molecular level of injecting selected vibrational energy into an ice. Our findings can be summarised as follows. 

\begin{itemize}
\item
We have shown that pASW goes through a restructuring process reorienting some of its molecules into a crystalline-like structure upon resonant absorption of IR radiation. Experimental results are analysed via synthetic spectra of pASW and Ic, as well as by fitting individual oscillators. Theoretical simulated ices give similar results, both in terms of the spectral signatures upon irradiation and the modification of identified oscillators in the ice.

\item
We have shown that the crystallisation-like phenomenon can be successfully simulated using MD and is hence attributed to localised heating of the ice caused by vibrational excitation, resulting in a greater proportion of H-bonding in the ice after each laser pulse.

\item
This phenomenon saturates within five minutes of FEL irradiation, modifying upwards of 94~\% of the irradiated ice.

\item
Future work includes the upgrade of the system to include a quadrupole mass spectrometer to directly monitor IR-induced desorption in the gas phase and to better constrain other astrochemical, surface science and catalysis relevant phenomena such as IR-induced surface diffusion, segregation and reaction under ultrahigh vacuum conditions. A dedicated study on the irradiation of the THz modes of pASW is also underway.
\end{itemize}

\begin{acknowledgement}


The authors thank the FELIX Laboratory team for their experimental assistance and scientific support. Furthermore, we would like to thank Dr. Lex van der Meer for the useful discussions during the preparation of the manuscript. The LISA UHV setup was designed, constructed and managed at the FELIX Laboratory by the group of Dr. Ioppolo. This work was supported by the Royal Society University Research Fellowship (UF130409); the Royal Society Research Fellow Enhancement Award (RGF/EA/180306); and the Royal Society Research Grant (RSG/R1/180418). Travel support was granted by the UK Engineering and Physical Sciences Research Council (UK EPSRC Grant EP/R007926/1 - FLUENCE: Felix Light for the UK: Exploiting Novel Characteristics and Expertise); the LASERLAB-EUROPE support (grant agreement no. 654148, European Union's Horizon 2020 research and innovation programme); and Short Term Scientific Missions (COST Actions CM1401 and TD1308).

\end{acknowledgement}




\clearpage

\section{Supplementary Information}

\renewcommand\thefigure{S\arabic{figure}} 
\renewcommand\thetable{S\arabic{table}} 
\setcounter{figure}{0} 
\setcounter{table}{0} 

\section*{Details of the experimental setup}

\begin{figure}[ht]
\centering
  \includegraphics[width=\columnwidth]{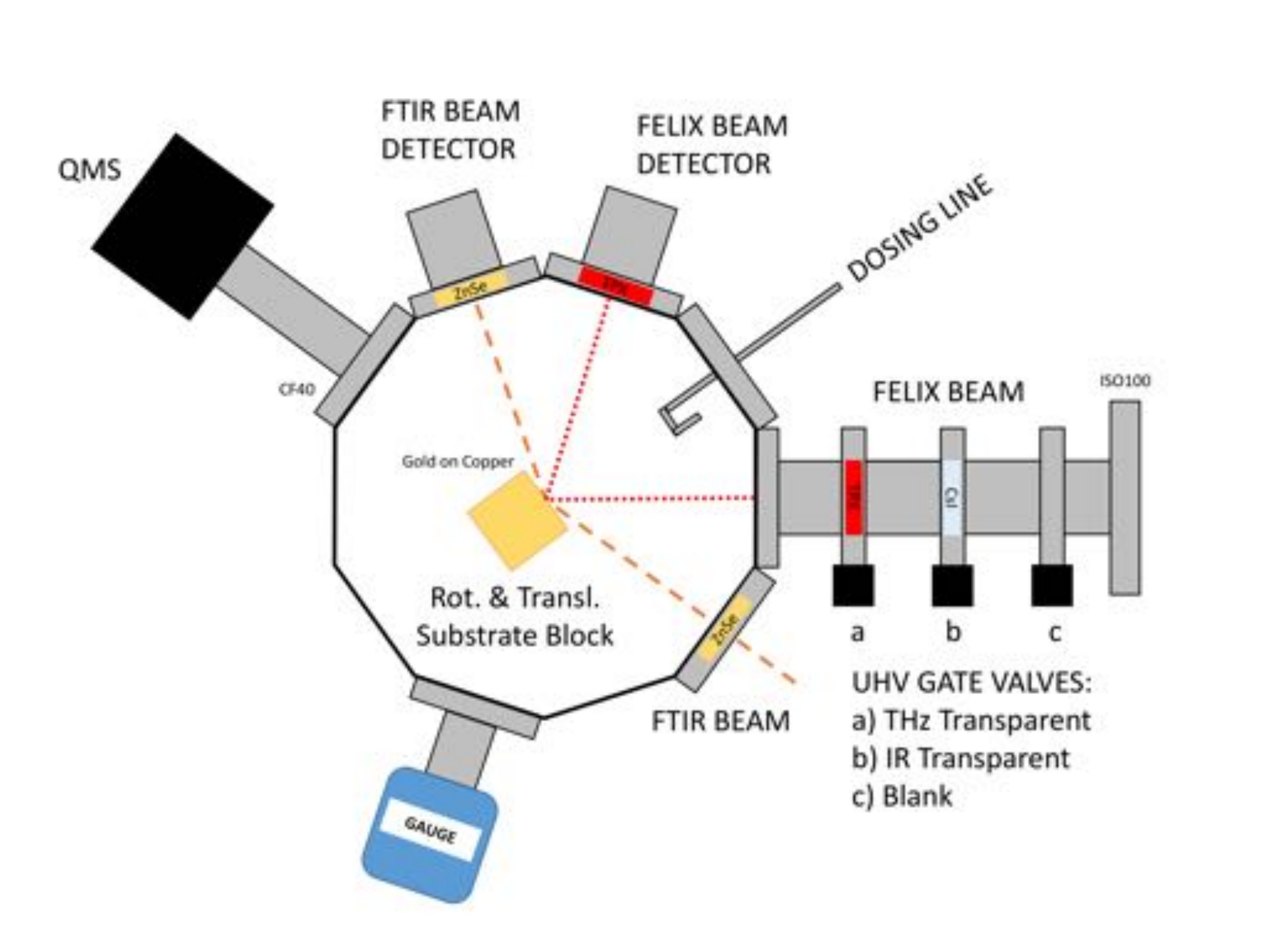}
    \caption{Schematic top view of the Laboratory Ice Surface Astrophysics (LISA) end-station at the FELIX Laboratory used to perform infrared-resonant vibrational-induced restructuring of space related ice samples.}
  \label{fig:UHV_setup}
\end{figure}

The Laboratory Ice Surface Astrophysics (LISA) end-station at the FELIX Laboratory is an ultrahigh vacuum (UHV) setup developed, constructed and managed by the group of Dr. Ioppolo based at QMUL in the UK. The system is designed and optimised to perform selective IR/THz irradiation of solid phase molecules when coupled to the tunable, high-power and short-pulsed light from the free-electron lasers (FELs) FELIX-1 ($\sim$~30-150~$\mu$m) and FELIX-2 ($\sim$~3-45~$\mu$m). Selective IR/THz experiments at a FEL facility are needed to better understand the nature of IR/THz modes in solids; time-resolved dynamics and energy relaxation within condense matter; and fundamental processes like diffusion, segregation, reaction and ultimately desorption of molecules at the surface and in the bulk of ice layers. Thus the system finds applications in astrochemistry, astrobiology, surface science and catalysis. Figure~\ref{fig:UHV_setup} shows a schematic of the setup configuration used at the time of the present work. Briefly, the main UHV chamber, a 8.0 inches spherical decagon from Kimball Physics with top and bottom DN160CF and other ten DN40CF flanges at its sides, is connected at its bottom to a DN100CF turbo molecular TURBOVAC 151 pump (Oerlikon Leybold Vacuum) through a DN160-100CF adapter. On top of the DN160CF flange of the main chamber there is a heavy duty $z$-translator with a stroke of 50.8 mm (LewVac) connected to a custom-made 15 mm off-axis DN160-63CF adapter (LewVac), a two stage differentially pumped DN63CF rotary platform (LewVac), a custom-made DN100-63CF adapter (LewVac), and a closed-cycle helium CH-204SB cryostat head with a HC-4E water-cooled compressor (Sumitomo).

The cryostat head is in thermal contact with a custom-made 30$\times$30$\times$50 mm (l$\times$w$\times$h) oxygen-free high thermal conductivity (OFHC) copper block substrate with four optically flat faces all coated in gold to ensure maximal reflection in the IR-THz spectral range. In this experimental configuration, the selected substrate face exposed to the FELIX beam is centred in the main chamber due to the off-axis flange adapter and it can be moved in height by means of the $z$-translator to allow irradiation of the gold-coated surface at different spots of the substrate face. The $z$-translation stage is also used to remove the substrate from the line of sight of the FELIX beam such that a direct measurement and alignment of the beam is possible at any time through the demountable ZnSe (Crystran) viewport (LewVac) at the back of the chamber, opposite the entrance of the FEL light.

The substrate temperature is controlled in the range 15-300~K by means of a Kapton tape heater (Omega) connected to the OFHC copper block and regulated with a model 335 cryogenic temperature controller (LakeShore) that reads temperatures through an uncalibrated DT-670B-SD silicon diode (LakeShore) with a CO mount adapter fixed at the bottom of the substrate. The block has a 3 mm hole through to allow for the wiring of the silicon diode. Pressure in the main chamber is monitored by means of a wide range gauge (1$\times$10$^3$ -- 1$\times$10$^{-9}$~mbar; Edwards Vacuum) coupled to a TIC Instrument Controller 3 (Edwards Vacuum). Active Pirani vacuum gauges are connected in various parts of the system, \emph{e.g.} between the turbo pump and the SH-110 dry scroll pump (Agilent) and in the dosing line to measure its base pressure. 

A 6~mm Swagelok tubing dosing line is used to prepare pure gases and gas mixtures prior to their dose inside the main chamber. An active strain gauge (2000~mbar; Edwards Vacuum) is used to control gases mass-independently during gas mixture preparation. Gases are then introduced in the main chamber through an all-metal leak valve (LewVac) connected to a custom-made double-sided DN40CF flange (LewVac) with a 6~mm central tube facing the walls of the chamber to allow for a background deposition of gases onto the substrate, \emph{i.e.} molecules are deposited at random angles with respect to the substrate surface to better simulate deposition in space on interstellar dust grains. A background deposition also allows for a more uniform ice deposition all around the OFHC copper block. Hence once a selected ice is deposited, FELIX irradiation can be carried-out at multiple unirradiated spots of the same deposited ice (\emph{i.e.} 6 unprocessed spots per block face) maximising the use of the beamtime shift (\emph{i.e.} 8 hours per shift at the FELIX Laboratory) and allowing for more systematic and reproducible studies. 

Some of the DN40CF flanges of the main chamber are blank, others as shown in Figure~\ref{fig:UHV_setup} mount demountable ZnSe (Crystran) and TPX (Tydex) viewports (LewVac) to perform infrared spectroscopy and measure the FEL beam power after the sample. Moreover two UHV DN40CF gate valves with CsI (Crystran) and TPX windows, respectively, are mounted in series together with a blank DN40CF gate valve in the FEL beam path before the chamber to allow for a faster change between FELIX-1 and -2 during a single shift with no need for opening the vacuum line. The use of these valves substantially lower the water contamination in the UHV chamber, which is not exposed to the vacuum pressure of the FEL line ($\sim$~10$^{-6}$ mbar at the end-station connection) at any time. An external liquid nitrogen cooled mercury cadmium telluride (MCT) detector is used to acquire mid-IR data by means of the Fourier-Transform Infrared spectrometer in reflection-absorption mode (Vertex 80v from Bruker). While the spectrometer is pumped with a dry scroll pump (Edwards Vacuum), external optics and the MCT detector are in purge boxes filled with pure nitrogen gas. Finally, Figure~\ref{fig:UHV_setup} shows a residual gas analyser mass spectrometer (MKS Instruments) not used in this work.

\section*{Full IR spectra of experimentally irradiated pASW and FEL power dependence}

\begin{figure}[ht]
\centering
  \includegraphics[width=\columnwidth]{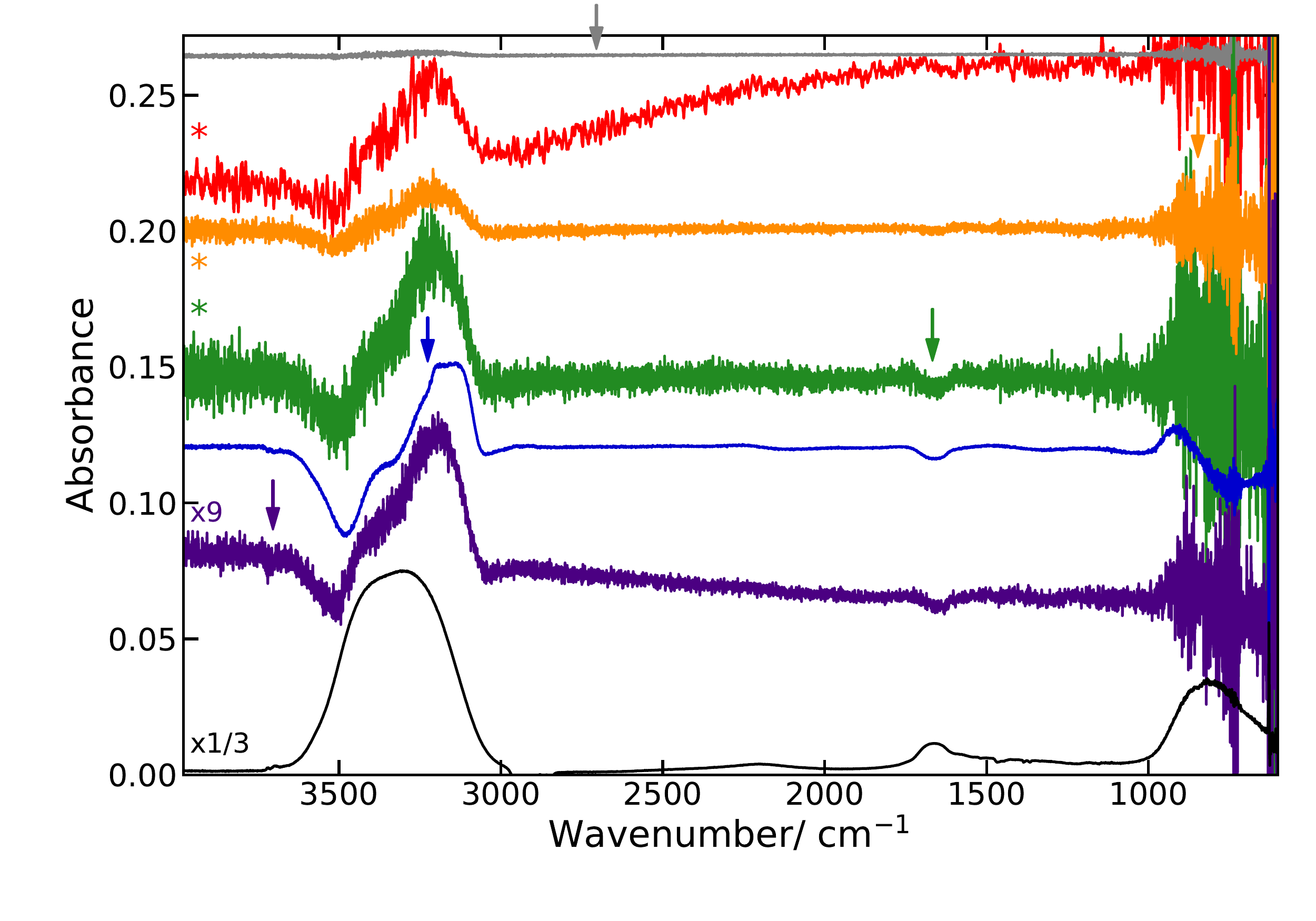}
    \caption{Experimental spectra of pASW before and after irradiation. The pASW signal before irradiation (black line) is divided by three to aid comparison with the difference spectra after irradiation at 2.7~$\mu$m (dangling OH stretching mode, violet) 3.1~$\mu$m (bulk stretching mode, blue), 6.0~$\mu$m (bending mode, green), 11.8~$\mu$m (libration mode, orange), 37.5~$\mu$m (interlayer, red), and 3.7~$\mu$m (off-resonance, gray) irradiations. Difference spectra of irradiations at the bending, libration, and THz modes are corrected for the band strengths of the irradiated modes (*)\cite{Hudgins:1993,Bouilloud:2015}. The irradiation at 37.5~$\mu$m was carried out without attenuation (100~mJ/macropulse) and is smoothed due to low signal-to-noise. No band strength is available for the surface modes at 2.7~$\mu$m, and thus for this irradiation (violet), the spectrum was multiplied to match the depth of the abundance change after irradiation at 3.1~$\mu$m.}
  \label{fig:fullspec}
\end{figure}

Figure~\ref{fig:fullspec} is the spectral extension of Figure~2. Therefore all details are discussed in the main text. Additionally, here we present a difference spectrum of an irradiation in the Far InfraRed/TeraHetz (FIR/THz; 37.5~$\mu$m / $\sim$8 THz) range to test the irreversible effects of exposing intermolecular (\emph{i.e.} interlayer) modes to a FEL beam. The difference spectrum of the irradiation in the THz mode is corrected for its corresponding band strength (4.8~$\times$~10$^{-18}$~cm\,molec$^{-1}$) to allow for a direct visual comparison to the stretching mode irradiation (band strength 2.0~$\times$~10$^{-16}$~cm\,molec$^{-1}$).\cite{Hudgins:1993} It should be noted that, during the experiment shown in the figure, the laser macropulse energy at all wavelengths, except that at 37.5~$\mu$m, was attenuated to match that in the 3~$\mu$m region ($\sim$~10~mJ/macropulse). Additionally, the THz spectrum (in red) is smoothed due to low signal-to-noise, despite this irradiation being performed with a FEL energy of $\sim$~100~mJ/macropulse. This irradiation represents the limit in the signal-to-noise during these experiments, due to the difficulty of extracting the irradiation-modified signal from the small amount of background deposition in the ultrahigh vacuum chamber. An upgraded experimental system with a lower base pressure will remedy this in future investigations, providing a higher signal-to-noise limit. Moreover, it should be noted that the irradiation at 37.5~$\mu$m sits on the wing of a strong and broad THz mode of pASW with a peak at $\sim$~48~$\mu$m, \emph{i.e.} beyond the spectral range of FELIX-2.\cite{Moore:1992} The choice of irradiating at 37.5~$\mu$m was dictated by the transmission curve ($>$85~\% between 0.6--38~$\mu$m) of the cesium iodide (CsI) window used to separate the FELIX beam line from the LISA setup. In a future study we will investigated the THz spectral range accessible by FELIX-1 ($\sim$~30--150~$\mu$m) in combination with a Polymethylpentene (TPX) window with a transmission curve $>$50~\% between 30--3000~$\mu$m (10--0.1~THz).

\begin{figure}[ht]
\centering
  \includegraphics[width=\columnwidth]{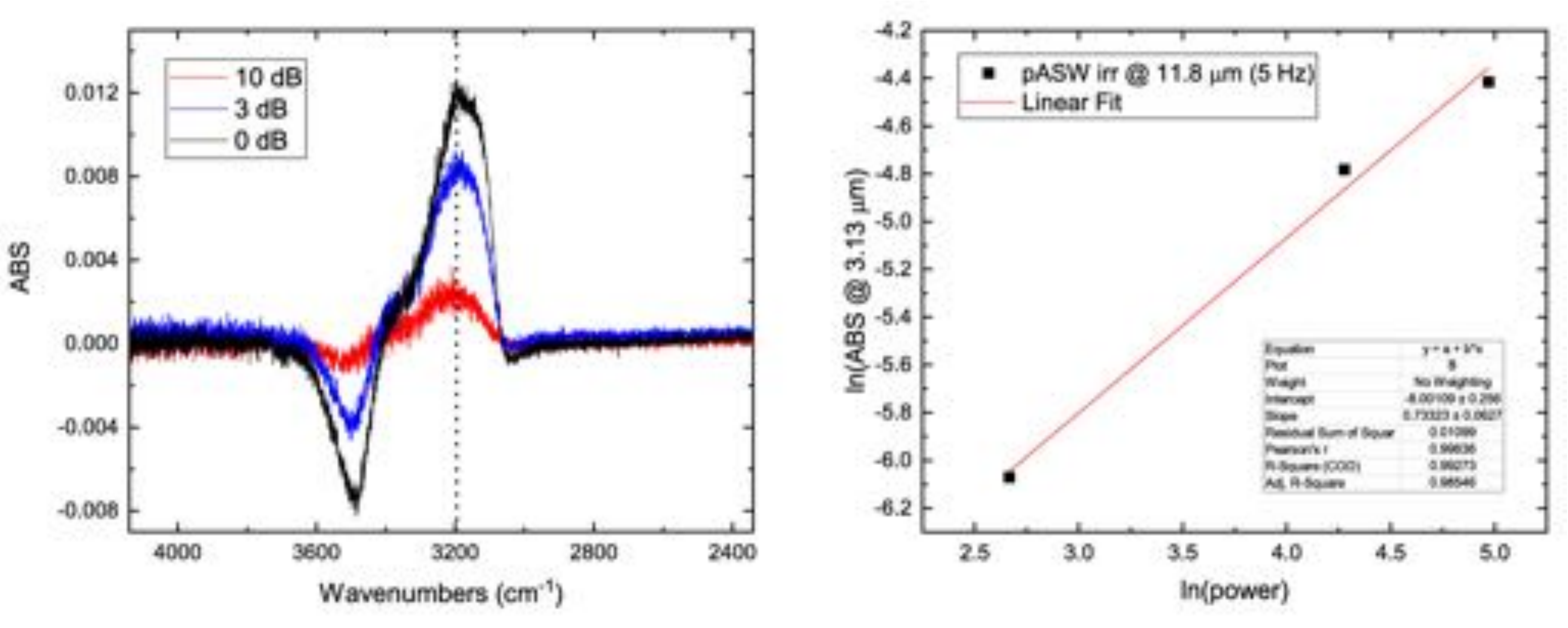}
    \caption{Left panel: RAIR difference spectra of pASW in the  3~$\mu$m region before any irradiation and after three consecutive irradiations at 11.8~$\mu$m with 10 decibels (dB, red), 3 dB (blue) and 0 dB (black) power attenuation, respectively. The dotted vertical line at 3.13~$\mu$m indicates the peak position for the absorption band values used in the right panel. Right panel: Absorption band values at 3.13~$\mu$m plotted versus the FEL power for each irradiation. The red line is the best linear fit to the data, which presents a slope of 0.73 +/- 0.06.}
  \label{fig:single_photon_process}
\end{figure}

\section*{Synthetic spectra generated from optical constants}

The optical properties of a thin film can be treated using a three phase vacuum-absorber-metal system model, where the Fresnel coefficients for phases 0, 1, and 2 in the parallel and perpendicular directions are:

\begin{equation}
    \begin{aligned}
        r_{||012} = \frac{r_{||01} + r_{||12}e^{-2i\beta}}{1 + r_{||01}r_{||12}e^{-2i\beta}} \\
        r_{\perp012} = \frac{r_{\perp01} + r_{\perp12}e^{-2i\beta}}{1 + r_{\perp01}r_{\perp12}e^{-2i\beta}}
    \end{aligned}
\end{equation}

In this formalism, the individual Fresnel coefficients are:
\begin{equation}
    \begin{aligned}
        r_{||01} &= \frac{n_1 \cos\theta - \sqrt{1- (\frac{\sin\theta}{n_1})^2}}{n_1 \cos\theta + \sqrt{1- (\frac{\sin\theta}{n_1})^2}} \\
        r_{||12} &= \frac{n_2 \sqrt{1- (\frac{\sin\theta}{n_1})^2} - n_1 \sqrt{1- (\frac{\sin\theta}{n_2})^2}}{n_2 \sqrt{1- (\frac{\sin\theta}{n_1})^2} + n_1 \sqrt{1- (\frac{\sin\theta}{n_2})^2}}\\
        r_{\perp01} &= \frac{\sqrt{1- (\frac{\sin\theta}{n_1})^2} - n_1 \cos\theta}{\sqrt{1- (\frac{\sin\theta}{n_1})^2} + n_1 \cos\theta} \\
        r_{\perp12} &= \frac{n_1 \sqrt{1- (\frac{\sin\theta}{n_1})^2} - n_2 \sqrt{1- (\frac{\sin\theta}{n_2})^2}}{n_1 \sqrt{1- (\frac{\sin\theta}{n_1})^2} + n_2 \sqrt{1- (\frac{\sin\theta}{n_2})^2}}
    \end{aligned}   
\end{equation}

and $\beta$ describes the phase change as:
\begin{equation}
\beta = \frac{2\pi n_1d\cos\theta}{\lambda}
\end{equation}
where n$_p$ is the optical constant of phase p, d is the film thickness, $\lambda$ is the wavelength, and $\theta$ is the angle of incidence.\cite{Heavens:1991,Horn:1995,Teolis:2007}
Synthetic spectra were generated based on available optical constants for samples of pASW deposited at 15~K,\cite{Mastrapa:2009} Ic deposited at 150~K and measured at 17~K,\cite{Mastrapa:2009} and the gold substrate.\cite{Ordal:1987} The RAIRS angle, $\theta$, was 17.9~$\pm$~1.2$^\circ$, with the final spectrum for each ice sample generated from spectra synthesised every 0.25$^\circ$ convolved with a Gaussian profile to take into account the experimental beam profile. Spectra were synthesised for ice samples of thicknesses up to 0.3~$\mu$m at steps of 0.01~$\mu$m.

\begin{figure}[ht]
\centering
  \includegraphics[width=\columnwidth]{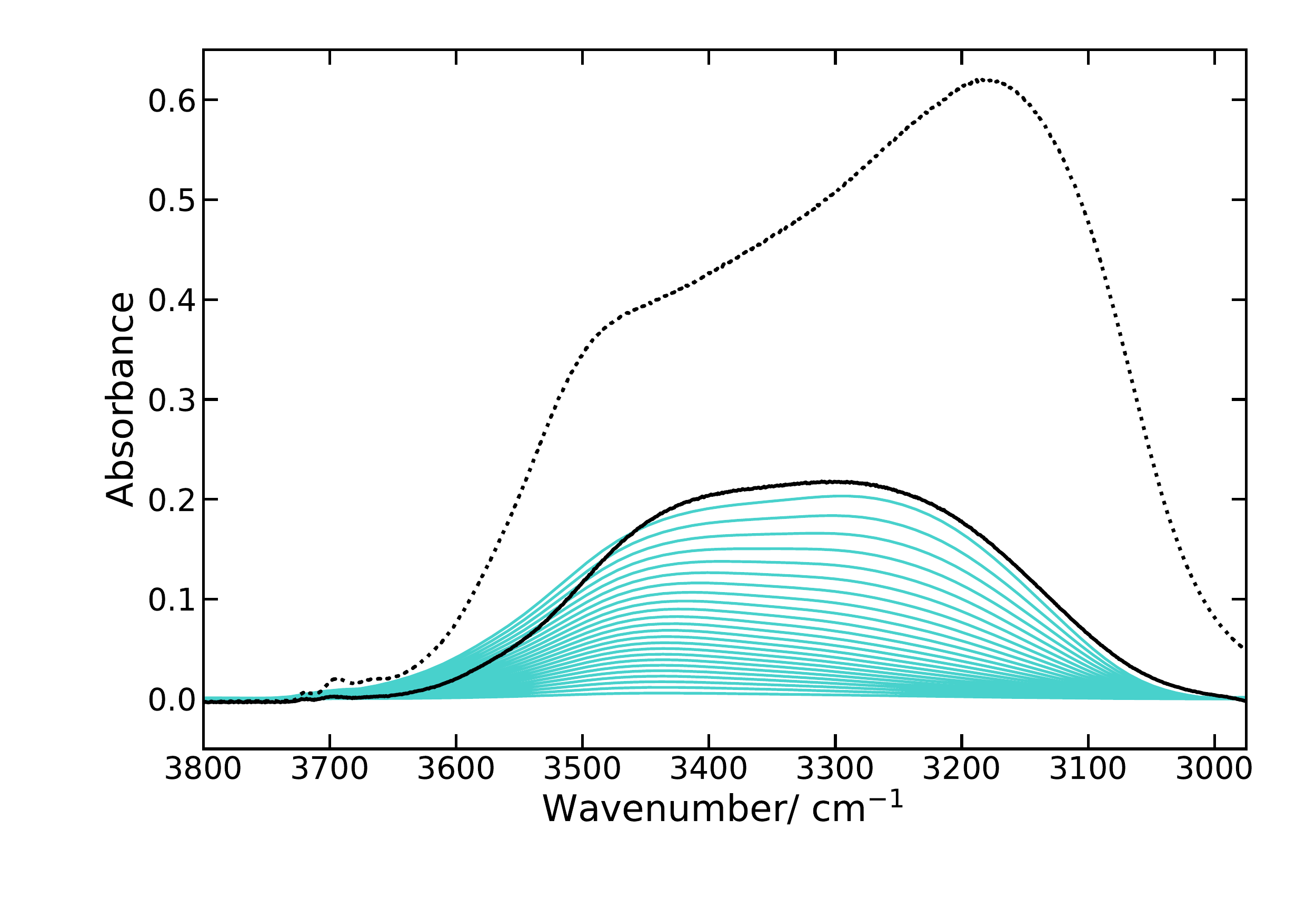}
  \includegraphics[width=\columnwidth]{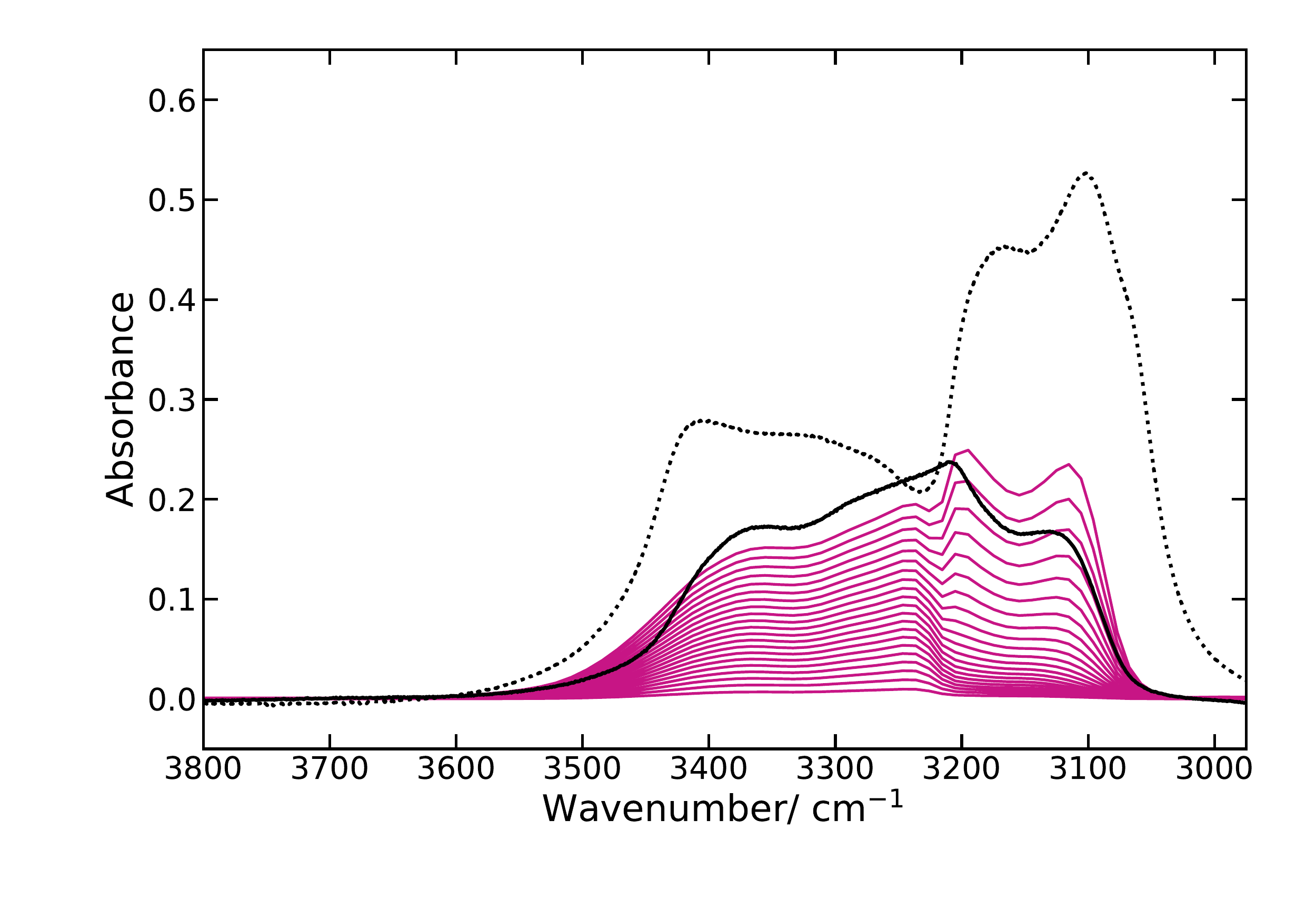}
    \caption{Synthetic spectra generated for pASW (upper panel) and Ic (lower panel). In the upper panel, experimental pASW spectra for ices deposited at 17~K (black) are plotted for two thicknesses, with synthetic spectra for pASW of thicknesses 0.01--0.24~$\mu$m (cyan). In the lower panel, experimental Ic spectra for ices deposited at 150~K (black) are plotted for two thicknesses, with synthetic spectra for Ic of thicknesses 0.01--0.21~$\mu$m (magenta).}
  \label{fig:synth}
\end{figure}

Figure~\ref{fig:synth} presents the experimental spectra of the pASW discussed in the main text as well as a thicker pASW sample prepared and measured using the same experimental setup (dotted line). The thickness derived from calculating the area under the peak of the experimental spectrum of the thinner pASW sample is $\sim$~0.3~$\mu$m (assuming a density of 0.78~g\,cm$^{-3}$ and a band strength of 2~$\times$~10$^{-16}$~cm\,molec$^{-1}$).\cite{Mastrapa:2009} Overlaid on the experimental spectra are simulated spectra for ices from 0.01 -- 0.24~$\mu$m thickness. The closest matching simulated spectrum is 0.24~$\mu$m thick. These values are also consistent with the expected deposition thickness as derived from deposition temperature and pressure conditions. It should be noted, however, that, in this case, the exact number of molecules deposited on the surface cannot be derived directly from the pressure reading, as we do not know the extent of the cryopumping in the system which reduces the pressure near the surface.

As for pASW, experimental and synthetic spectra of crystalline Ic are presented in Figure~\ref{fig:synth}. The closest matching simulated spectrum to the thinner Ic experimental spectrum is 0.21~$\mu$m thick. The Ic irradiation data presented in Figure~3 (lower panel) is based on the thicker Ic deposition (dotted line, Figure~\ref{fig:synth} lower panel).

\section*{Oscillator fitting method}
\begin{figure}[ht]
\centering
  \includegraphics[width=\columnwidth]{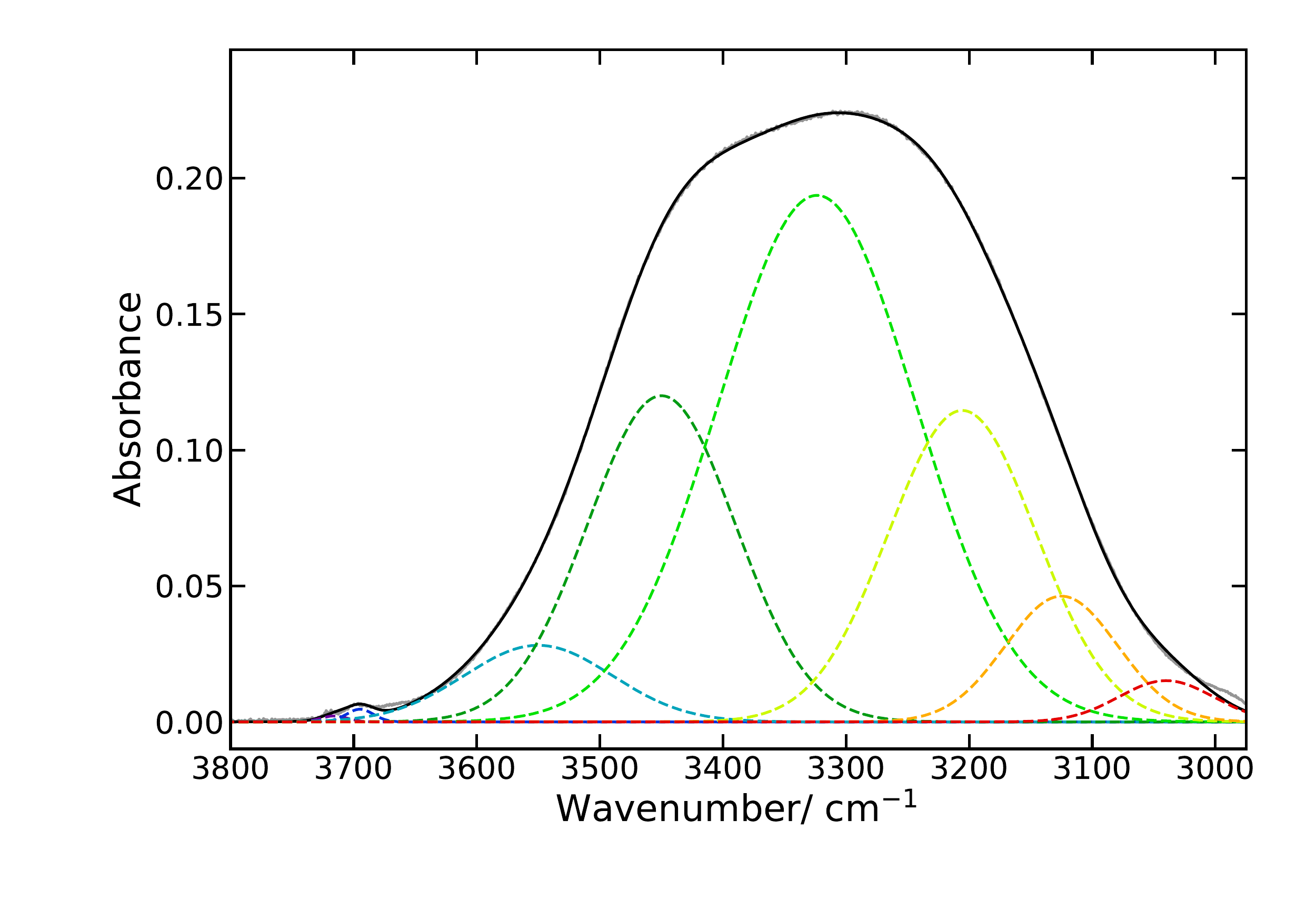}
  \includegraphics[width=\columnwidth]{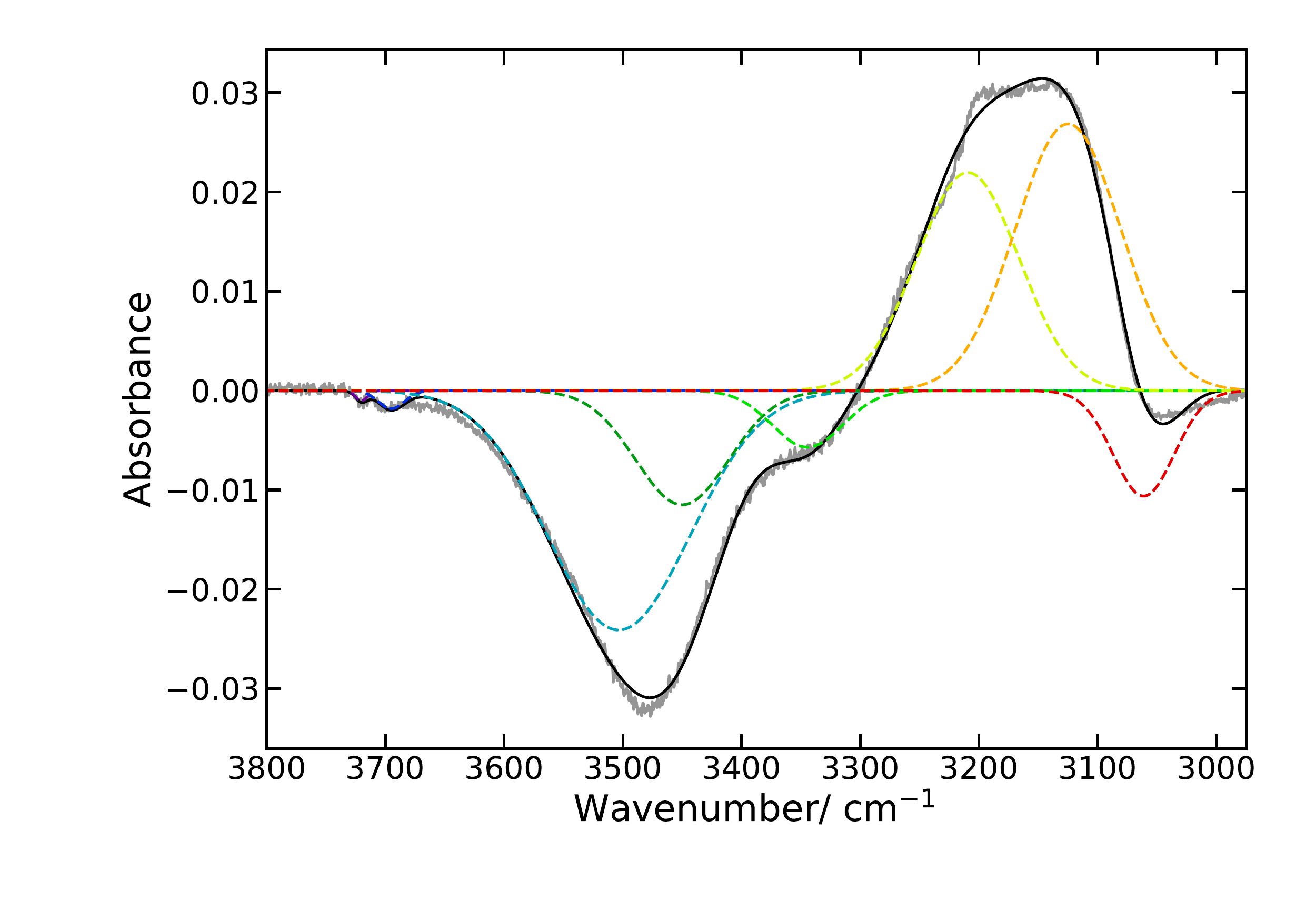}
  \caption{Fitting process for original deposited spectra (upper panel) and difference spectra (lower panel). Original spectrum (gray) is fitted with a combination of eight Gaussian functions (coloured) to produce the final best fit spectrum (black).}
  \label{fig:oscillator_fitting}
\end{figure}

As described in the main text, oscillator classes are labelled depending on their bonding order within the ice network, as follows: D stands for ``donor'' and A for ``acceptor''. The least bound molecules at the surface are labelled DA, followed by the triply bound species DAA and DDA, while the fully tetrahedrally bound species in the bulk of the ice are denoted DDAA. The oscillator fitting method fits these various oscillator classes for a wide range of ice structures and thicknesses.
The oscillator fitting method for experimental spectra is illustrated in Figure~\ref{fig:oscillator_fitting}. A combination of eight Gaussian functions (G1 -- G8) is fitted to both original ice spectra (upper panel) and to difference spectra (lower panel), confirming the applicability of the chosen methodology to the analysis of these data. The attribution of oscillators is as follows: DA, G1; DAA, G2; DDA, G3; and DDAA G6+G7. The identity of the other oscillator families has not yet been unequivocally determined in the literature.

\section*{Possible missing quantum effects in MD simulations}

In the present paper, the experimental findings have been explained by classical effects. Off course, a proton-rich molecule like water will show many quantum effects at low temperature. We know, for instance, that the calculated radial distribution function and the IR spectrum change when taking into account quantum effects. To test this, we have performed Ring-Polymer-Contraction Molecular Dynamics simulations, following \citep{Markland:2008} and using the i-pi package in \textsc{Lammps}\citep{Ceriotti:2014}. The intramolecular interaction potential has been changed to the q-TIP4P/F \citep{Habershon:2009} potential. The system was simulated at 30~K to reduce the number of required beads. In this configuration, 200 beads for the intramolecular interactions were applied and 50 beads for the long range effects. The individual trajectories of the beads show indeed a strong delocalisation of the hydrogen atoms. As a result the RDF smoothened and broadened. The same holds for the calculated IR spectrum (both not shown). Exposing the individual beads to an external electric field, results in the same qualitative behaviour. This indicates that the energy that is put in the system by irradiation is sufficiently high to overcome specific quantum effects. Moreover, the experimental findings that irradiation leads to structural rearrangement is in further support of this being a classical effect where, for instance, tunneling is not expected to play a significant role. Finally, another aspect of the classical nature of the simulations is that the excitation is not quantized and that there is no clear threshold between a one-photon or two-photon regime. 

\providecommand{\latin}[1]{#1}
\makeatletter
\providecommand{\doi}
  {\begingroup\let\do\@makeother\dospecials
  \catcode`\{=1 \catcode`\}=2 \doi@aux}
\providecommand{\doi@aux}[1]{\endgroup\texttt{#1}}
\makeatother
\providecommand*\mcitethebibliography{\thebibliography}
\csname @ifundefined\endcsname{endmcitethebibliography}
  {\let\endmcitethebibliography\endthebibliography}{}

\end{document}